\documentclass[11pt]{article}

\usepackage[utf8]{inputenc}
\usepackage{amsmath}
\usepackage{amssymb}
\input{epsf}
\usepackage{natbib}
\usepackage[margin=1in]{geometry}

\usepackage[raggedright]{titlesec}
\usepackage[hang,flushmargin]{footmisc} 
\usepackage{graphicx}
\usepackage{framed}
\usepackage{color}
\usepackage{yfonts}
\usepackage[dvipsnames,table]{xcolor}
\usepackage{pifont}
\usepackage{amsmath,amssymb,mathtools}
\usepackage{multirow}
\usepackage{amsfonts}
\usepackage{caption}
\usepackage{subcaption}
\usepackage{float}
\usepackage{threeparttable}
\usepackage{esvect}
\usepackage{pgf}
\usepackage{nicematrix}
\usepackage{tikz}
\usepackage{tikz-cd}
\usepackage{fancyvrb}
\usetikzlibrary{positioning}
\usepackage{bbm}
\usepackage{colortbl}
\usepackage{pdflscape}
\usepackage{booktabs}
\usepackage{amsmath}
\usepackage{mathdots}
\usepackage{yhmath}
\usepackage{cancel}
\usepackage{siunitx}
\usepackage{array}
\usepackage{multirow}
\usepackage{gensymb}
\usepackage{tabularx}
\usepackage{booktabs}
\usepackage{tabstackengine}
\usetikzlibrary{fadings}
\usetikzlibrary{patterns}
\usetikzlibrary{shadows.blur}
\usetikzlibrary{shapes}
\usepackage{placeins}
\usepackage{comment}
\usepackage[ruled,linesnumbered,commentsnumbered]{algorithm2e}
\SetKwComment{Comment}{// }{}

\usepackage[font=small,labelfont=bf]{caption}

\usepackage{hyperref}
\hypersetup{
    colorlinks=true,
    linkcolor=NavyBlue,
    filecolor=magenta,      
    urlcolor=cyan,
    citecolor=MidnightBlue
}
 
\usepackage{rotating}
\usepackage{ntheorem}
\makeatletter
\renewtheoremstyle{plain} 
{\item{\theorem@headerfont ##1\ ##2\theorem@separator}~}
{\item{\theorem@headerfont ##1\ ##2\ (##3)\theorem@separator}~}
\makeatother

\theoremheaderfont{\normalfont\bfseries}

\usepackage{enumitem}
\newlist{Step}{enumerate}{2}
\setlist[Step]{label={{Step \arabic*.}}, leftmargin=*}
\usepackage{setspace}

\makeatletter
\newcommand\circled[1]{%
  \mathpalette\@circled{#1}%
}
\newcommand\@circled[2]{%
  \tikz[baseline=(math.base)] \node[draw,circle,inner sep=2pt] (math) {$\m@th#1#2$};%
}
\makeatother

\makeatletter
\newcommand\circledblue[1]{%
  \mathpalette\@circledblue{#1}%
}
\newcommand\@circledblue[2]{%
  \tikz[baseline=(math.base)] \node[draw,circle, fill=blue!20, inner sep=2pt] (math) {$\m@th#1#2$};%
 }

\renewenvironment{abstract}
 {\begin{center}\normalsize\textsc{Abstract}%
 \end{center}\begin{quote}\normalsize}
 {\end{quote}}

\definecolor{lightgray}{gray}{0.9}

\usepackage{apptools}

\renewcommand{\Vec}[1]{\boldsymbol{#1}}
\newcommand{\Mat}[1]{\mathbf{#1}}

\newcommand{\bD}{\mathbf{D}}

\newcommand{\bU}{\mathbf{U}}

\newcommand{\bW}{\mathbf{W}}

\newcommand{\bZ}{\mathbf{Z}}

\newcommand{\R}{\mathbb{R}}

\DeclareMathOperator*{\argmin}{arg\,min}

\newcommand{\node}{\mathfrak{t}}

\definecolor{cyan}{cmyk}{1, 0.2, 0, 0} 

\definecolor{green}{cmyk}{1,0,1,0} 

\definecolor{darkred}{rgb}{0.545, 0, 0}

\makeatletter
\@ifclassloaded{article}{

}{}
\makeatother

\newcommand{\method}{CoMDS}
\newcommand{\localmethod}{LoCoMDS}
\newcommand{\metak}{Meta-Spec (kPCA)}
\newcommand{\metau}{Meta-Spec (UMAP)}
\newcommand{\msne}{Multi-SNE}

\title{\bf Consensus dimension reduction via multi-view learning}
\author{
    Bingxue An\\
    University of Notre Dame\\
    \texttt{ban2@nd.edu}
    \and
    Tiffany M. Tang\\
    University of Notre Dame\\
    \texttt{ttang4@nd.edu}
}

\date{}

\begin{document}

\maketitle

\begin{abstract}
A plethora of dimension reduction methods have been developed to visualize high-dimensional data in low dimensions. However, different dimension reduction methods often output different and possibly conflicting visualizations of the same data. This problem is further exacerbated by the choice of hyperparameters, which may substantially impact the resulting visualization. To obtain a more robust and trustworthy dimension reduction output, we advocate for a \textit{consensus} approach, which summarizes multiple visualizations into a single \textit{consensus} dimension reduction visualization. Here, we leverage ideas from multi-view learning in order to identify the patterns that are most stable or shared across the many different dimension reduction visualizations, or \textit{views}, and subsequently visualize this shared structure in a single low-dimensional plot. We demonstrate that this consensus visualization effectively identifies and preserves the shared low-dimensional data structure through both simulated and real-world case studies. We further highlight our method's robustness to the choice of dimension reduction method and hyperparameters---a highly-desirable property when working towards trustworthy and reproducible data science.
\end{abstract}

\section{Introduction}\label{sec:intro}

Dimension reduction methods are a fundamental class of techniques in data analysis, which aim to find a lower-dimensional representation of higher-dimensional data while preserving as much of the original information as possible. 
These methods are extensively used in practice, including in exploratory data analyses to visualize data---arguably, one of the first and most vital steps in any data analysis \citep{Ray}.
Notably, in genomics, dimension reduction methods are ubiquitously applied to visualize high-dimensional single-cell RNA sequencing data in two dimensions \citep{Becht}. 
Beyond visualization, dimension reduction methods are also frequently employed to mitigate the curse of dimensionality \citep{Bellman}, engineer new features to improve downstream tasks like prediction \citep[e.g.,][]{massy1965principal}, and enable scientific discovery in unsupervised learning settings \citep{chang2025unsupervised}.
For example, many researchers have used dimension reduction in conjunction with clustering to discover new cell types and cell states \citep{WuSunny}, new cancer subtypes
\citep{Northcott}, and other substantively-meaningful structure in a variety of domains \citep{Beigen, Traven}.

Given the widespread use and need for dimension reduction methods, numerous dimension reduction techniques have been developed. Popular techniques include but are not limited to principal component analysis (PCA) \citep{Pearson-PCA, Hotelling-PCA}, multidimensional scaling (MDS) \citep{Torgerson-MDS, Kruskal-iMDS}, Isomap \citep{Tenenbaum-isomap}, locally linear embedding (LLE) \citep{Roweis-LLE}, t-distributed stochastic neighbor embedding (t-SNE) \citep{vdMaaten-t-SNE}, and uniform manifold approximation and projection (UMAP) \citep{McInnes-UMAP}.
Impressively, each aforementioned method has been cited more than 10,000 times, demonstrating their broad adoption and high impact.

Despite their incredible popularity however, there remains a significant problem in practice. 
Different dimension reduction methods are designed to preserve different types of structure in the data (e.g., see \citet{lee2007nonlinear} for a review)
and hence often output different, possibly conflicting visualizations of the same data \citep[e.g.,][]{Wang, Tsai}.
This problem is exacerbated by hyperparameters, which may substantially impact the resulting visualization \citep{Kobak2021, Kobak2019, ma2025uncovering}. 
To further add to this concern, a large proportion of research articles---estimated to be $\sim 44\%$ by previous work \citep{Jeon}---offer no justification for their choice of dimension reduction method.

\begin{figure}
    \centering
    \includegraphics[width=0.97\linewidth]{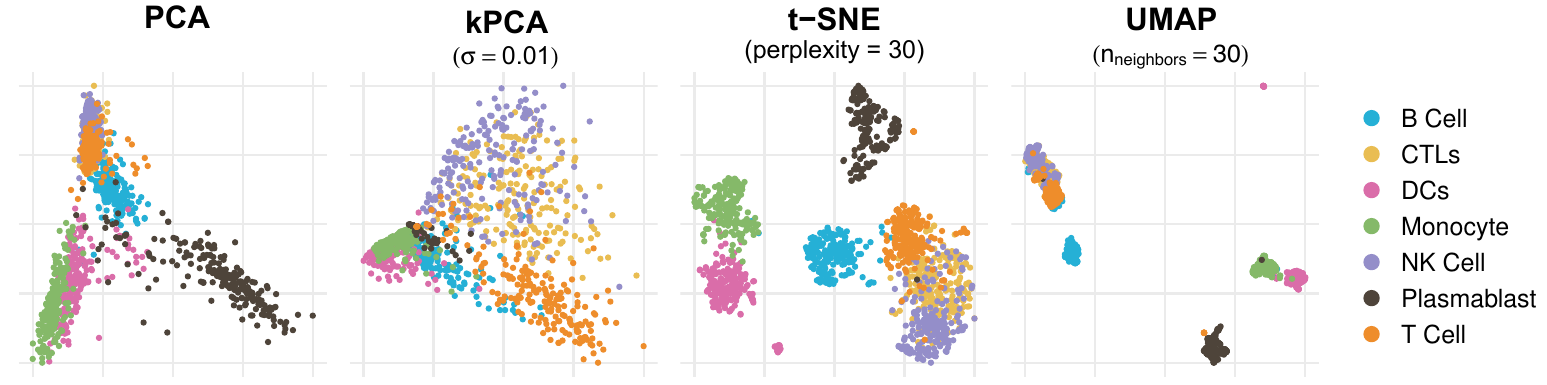}
    \caption{Comparison of different dimension reduction methods applied to single-cell RNA sequencing data from peripheral blood mononuclear cells, collected from individuals with HIV infection \citep{Kazer}, revealing visually different representations.}
    \label{fig:intro_example}
\end{figure}

We concretely demonstrate this heterogeneity of results in Figure~\ref{fig:intro_example} using single-cell RNA sequencing data from individuals with HIV infection \citep{Kazer}. In this motivating example, the four dimension reduction methods (PCA, kernel PCA (kPCA), t-SNE, and UMAP) yield markedly different visual representations of the same cells. 
While kPCA separates NK cells (purple), CTLs (yellow), and T cells (orange) more clearly than PCA, PCA better distinguishes between the monocytes (green) and plasmablasts (brown).
t-SNE and UMAP, on the other hand, produce tighter clusters, but may overstate their separation, as evidenced by the mixing of different cell types across distant clusters. For example, UMAP firmly places a lone plasmablast (brown) cell in the monocyte (green) cluster, which is not observed in t-SNE nor PCA.

A natural question thus arises for practitioners: which dimension reduction method should be trusted?
This is particularly challenging to answer in unsupervised learning settings, where there is no outcome to help guide model selection; put differently, the data points in Figure~\ref{fig:intro_example} are no longer labeled (or colored), making it near-impossible to reconcile and validate the different results.
Even more worrying is our natural human tendency to perceive patterns in what is actually randomly-generated data---a cognitive bias called the \textit{clustering illusion} \citep{gilovich2008we, falk1997making}.
Though heuristic metrics have been proposed to evaluate the quality of low-dimensional embeddings \citep{Huang-evaluation, Espadoto}, there still lacks a definitive answer on how to select the most appropriate dimension reduction method for a given dataset.

Thus, in lieu of selecting a single dimension reduction method, we propose to build upon a multi-view learning (or data fusion) paradigm, where we seek to combine information from multiple data views (in this case, dimension reduction methods) and extract the shared, or \textit{consensus}, signal across the different views (i.e., methods). 
The key idea draws from the stability principle \citep{yu2013stability,yu2020veridical,yu2024veridical}, which advocates that patterns consistently identified across multiple methods are more likely to be reliable and trustworthy, compared to idiosyncratic patterns that appear in only one method or one particular choice of hyperparameters.

However, though many multi-view dimension reduction methods exist \citep[e.g.,][]{singh2008relational, abdi2013multiple, lock2013joint, Zhang, tang2021integrated}, they were originally designed to combine different data modalities (e.g., text, image, and tabular data) and are not necessarily tailored to overcome the distinct challenges when integrating multiple dimension reduction outputs, which are inherently difficult to align due to their invariance to rotations, reflections, and scalings.
Meanwhile, \citet{Ma-Meta} recently introduced Meta-Spec, a spectral method for combining multiple dimension reduction visualizations into one. 
Meta-Spec cleverly circumvents the alignment issue by first converting the coordinate embeddings from each dimension reduction method into a normalized pairwise distance matrix, encoding the relative distance between each pair of data points in the low-dimensional representation. 
These normalized distance matrices are invariant to rotations, reflections, and scalings and hence can be more easily combined---and specifically in Meta-Spec, combined via a weighted average, with weights taken from spectral decompositions.
Researchers can then apply the dimension reduction method of their choosing to this averaged distance matrix to obtain their final Meta-Spec visualization. Nonetheless, though the flexibility to choose any dimension reduction method may be valuable in some scenarios, the final Meta-Spec visualization can also be highly sensitive to this choice, again raising the question of which dimension reduction method is most appropriate in unsupervised settings.

To address these existing challenges and improve the overall reliability of dimension reduction visualizations, we introduce Consensus Multidimensional Scaling (\method), a novel multi-view framework for combining multiple dimension reduction visualizations into a single, consensus dimension reduction visualization. 
Unlike existing methods, \method{} does not require the researcher to choose a dimension reduction method (beyond the input methods); instead, rooted in the stability principle, \method{} directly optimizes for the consensus embedding and finds the low-dimensional representation that best preserves the most stable, \textit{consensus} patterns across the different input dimension reduction methods. 
Furthermore, we demonstrate the flexibility of our framework by extending \method{} to a local variant, Local Consensus Multidimensional Scaling (\localmethod).
While \method{} is designed to preserve the overall \textit{global}, consensus patterns, \localmethod{} is more robust to outliers and aims to preserve the \textit{local}, consensus patterns across the different input methods.
Through a variety of simulated and real-world datasets, we demonstrate the effectiveness of \method{} and \localmethod{}, highlighting their ability to accurately extract stable, consensus patterns and yield more reliable and trustworthy dimension reduction visualizations.

The remainder of this work is organized as follows. In Section~\ref{sec:method}, we introduce our consensus dimension reduction framework and detail \method{}, followed by \localmethod{}. To help build intuition behind the proposed methods, we illustrate their performance on simulated examples, where we know the ground truth, in Section~\ref{sec:sims}. We then demonstrate the effectiveness of \method{} and \localmethod{} on numerous real-world datasets in Section~\ref{sec:case_studies}. In Section~\ref{sec:discussion}, we conclude with a discussion.

\section{A Framework for Consensus Dimension Reduction}\label{sec:method}
In this section, we introduce a multi-view framework for consensus dimension reduction, whereby multiple dimension reduction methods are applied to the same dataset and their shared or \textit{consensus} patterns are identified to produce a more trustworthy dimension reduction visualization.

Formally, suppose we have $M$ dimension reduction methods under consideration and data $\Mat{X}$ with $n$ samples and $p$ features. 
Let $\Mat{Z}^{(m)} \in \R^{n \times p_m}$ denote the low-dimensional embedding of $\Mat{X}$ obtained from the $m$-th dimension reduction method ($m = 1, \ldots, M$), and let $\Mat{Z}_i^{(m)} \in \R^{p_m}$ denote the embedding of the $i$-th data point in the $m$-th dimension reduction method ($i = 1, \ldots, n$).
Note that the $M$ dimension reduction embeddings could be from different dimension reduction methods (e.g., PCA, t-SNE, UMAP) and/or from the same dimension reduction method with different hyperparameter settings (e.g., t-SNE with different choices of perplexity). 
The overarching goal is to combine the $M$ dimension reduction embeddings $\Mat{Z}^{(1)}, \ldots, \Mat{Z}^{(M)}$ into a single consensus embedding $\hat{\Mat{Z}}^* \in \R^{n \times p^*}$ that captures the shared patterns across the many dimension reduction~outputs.

\begin{figure}
    \centering
    \includegraphics[width=1\linewidth]{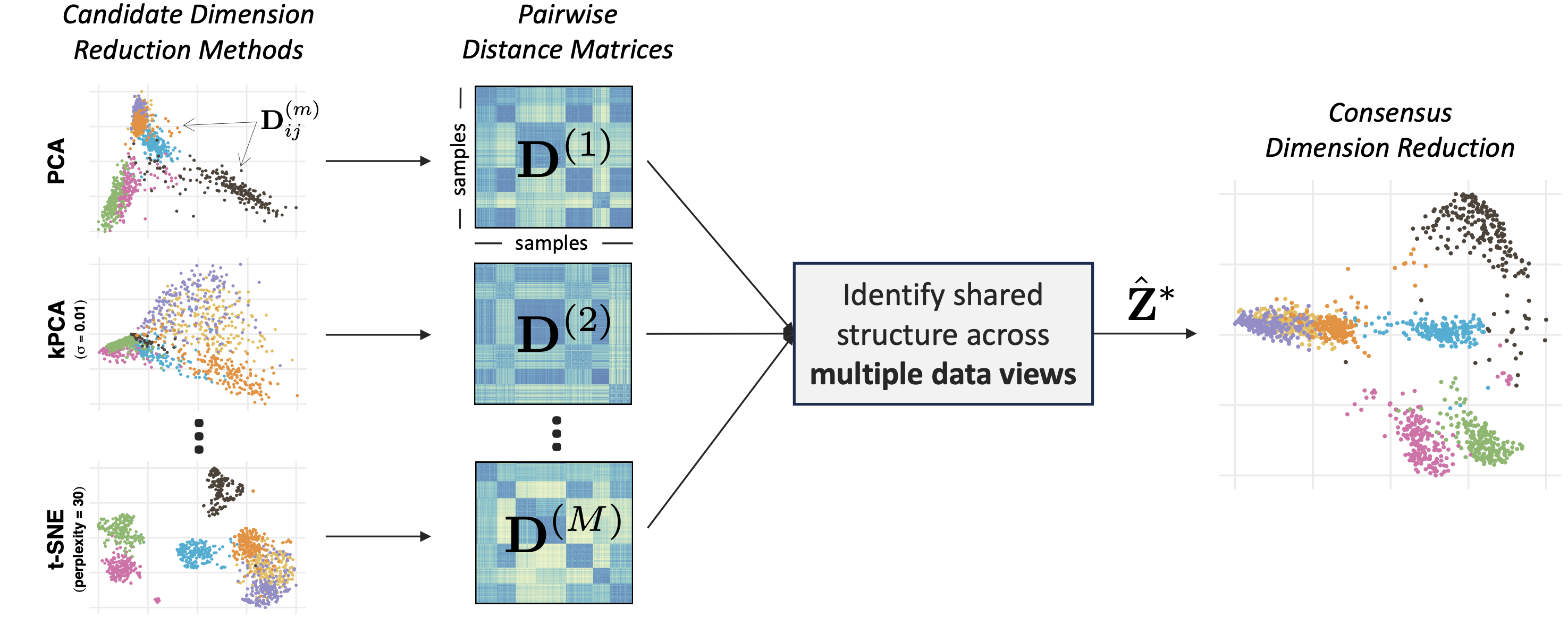}
    \caption{
        Overview of consensus dimension reduction framework. Given a set of candidate dimension reduction methods, we compute the pairwise distances in each dimension reduction space and subsequently extract the shared patterns across these distances to obtain our consensus dimension reduction embedding.
    }
    \label{fig:schema}
\end{figure}

To this end, we propose a multi-view framework for consensus dimension reduction (summarized in Figure~\ref{fig:schema}) and introduce Consensus Multidimensional Scaling (\method) and its localized variant, Local Consensus Multidimensional Scaling (\localmethod) next.

\subsection{Consensus Multidimensional Scaling (\method)}\label{subsec:comds}

Given input dimension reduction embeddings $\Mat{Z}^{(1)}, \ldots, \Mat{Z}^{(M)}$, Consensus Multidimensional Scaling (\method) produces a single consensus embedding $\hat{\Mat{Z}}^*$ via a simple two-step procedure.

\vspace{-0.5em}
\paragraph{Step 1: Encode dimension reduction embeddings as pairwise distance matrices.} 
For each dimension reduction method $m = 1, \ldots, M$, compute a pairwise distance matrix $\Mat{D}^{(m)} \in \R^{n \times n}$, encoding the distance between each pair of data points in the $m$-th dimension reduction representation. 
That is, $\Mat{D}^{(m)}_{ij} = d(\Mat{Z}_i^{(m)}, \Mat{Z}_j^{(m)})$ for each pair of data points $i, j = 1, \ldots, n$, where $d(\cdot, \cdot)$ is some user-defined distance metric (e.g., Euclidean distance).

\vspace{0.5em} 
Because different dimension reduction methods produce outputs that are generally defined on different coordinate systems and are invariant to rotations, reflections, and scalings, working directly with the raw dimension reduction embeddings $\Mat{Z}^{(m)}$ can be cumbersome.
Thus, similar to \citet{Ma-Meta}, the re-representation of dimension reduction embeddings as pairwise distance matrices (which are rotation- and reflection-invariant) is crucial to avoid complex alignment issues.

\vspace{0.5em} 
Next, while various multi-view dimension reduction methods could potentially be applied here to combine the different pairwise distance matrices $\Mat{D}^{(1)}, \ldots, \Mat{D}^{(M)}$, we specifically leverage a multi-view extension of multidimensional scaling (MDS) \citep{Torgerson-MDS, Torgerson1958} in order to exploit the inherent \textit{distance matrix} structure from Step 1. Recall, in the non-multi-view setting, (metric) MDS seeks to find a low-dimensional embedding $\hat{\Mat{Z}} \in \R^{n \times p^*}$ that best preserves the pairwise distances given in a single distance matrix $\Mat{D} \in \R^{n \times n}$ via:
\begin{align}
    \hat{\Mat{Z}} = \argmin_{\Mat{Z} \in \R^{n \times p^*}} \sum_{i < j} \Big\{ \underbrace{\Mat{D}_{ij}}_{\substack{\text{original}\\\text{distances}}} - \underbrace{d(\Mat{Z}_i, \; \Mat{Z}_j)}_{\substack{\text{distance in}\\\text{low-dim. space}}} \Big\}^2. \label{eq:mds}
\end{align}
We build upon MDS to identify the shared patterns across multiple distance matrices in Step 2.

\vspace{-0.5em}
\paragraph{Step 2: Extract shared patterns across multiple distance matrices.} 
Using the pairwise distance matrices $\Mat{D}^{(1)}, \ldots, \Mat{D}^{(M)}$ from Step 1, find the consensus embedding $\hat{\Mat{Z}}^* \in \R^{n \times p^*}$ that best preserves the pairwise distances observed in the input $M$ dimension reduction embeddings via a multi-view generalization of MDS:
\begin{align}
    \hat{\Mat{Z}}^*, \hat{\Mat{W}}^{(1)}, \ldots, \hat{\Mat{W}}^{(M)} = \argmin_{\substack{\Mat{Z} \in \R^{n \times p^*}\\ \Mat{W}^{(m)} \in \text{Diag}(p^*)}} \sum_{m = 1}^{M} \sum_{i < j} \Big\{ \underbrace{\Mat{D}^{(m)}_{ij}}_{\substack{\text{distance in } m^{th}\\\text{dimension reduction}\\\text{space}}} -\;\; \underbrace{ d(\Mat{W}^{(m)} \Mat{Z}_i, \; \Mat{W}^{(m)} \Mat{Z}_j)}_{\substack{\text{distance in consensus}\\\text{dimension reduction}\\\text{space}}} \Big\}^2, \label{eq:comds}
\end{align}
where $p^*$ is a pre-specified number of dimensions for the consensus output (typically, $2$ or $3$ for visualizations), $\hat{\Mat{W}}^{(1)}, \ldots, \hat{\Mat{W}}^{(M)}$ are $p^* \times p^*$ diagonal matrices that allows for different scalings of the distances in case $\Mat{D}^{(1)}, \ldots, \Mat{D}^{(M)}$ are measured on different scales, and $d(\cdot, \cdot)$ again is a distance measure (e.g., Euclidean distance).

\vspace{0.5em} 
Intuitively, \eqref{eq:comds} finds the low-dimensional representation $\hat{\Mat{Z}}^*$ such that the pairwise distances between points in the consensus space are as close as possible to the pairwise distances between points in each of the $M$ dimension reduction spaces.
We can thus think of the \method{} embedding $\hat{\Mat{Z}}^*$ as capturing the shared or consensus patterns across the different dimension reduction outputs.
Moreover, unlike Meta-Spec which applies a fixed (non-adaptive) normalization step to $\Mat{D}^{(1)}, \ldots, \Mat{D}^{(M)}$ and requires choosing a final dimension reduction method to obtain the consensus representation \citep{Ma-Meta}, \method{} adaptively learns the optimal scalings $\hat{\Mat{W}}^{(1)}, \ldots, \hat{\Mat{W}}^{(M)}$ for each distance matrix and directly optimizes for the consensus representation $\hat{\Mat{Z}}^*$ without the need to explicitly choose a final dimension reduction method---thus avoiding these arbitrary but consequential choices that are difficult to validate (see Sections~\ref{sec:sims} and \ref{sec:case_studies} for examples).

\paragraph{Connections to MDS literature.} 
While the application towards consensus dimension reduction is novel to this work, it is important to highlight that the optimization problem in \eqref{eq:comds} is not new, dating back to the 1970s in the psychometrics literature where it is referred to as Individual Differences in Multidimensional Scaling (INDSCAL) and primarily used to analyze dissimilarity data collected from multiple subjects \citep{Carroll1970}. 
Moreover, INDSCAL can be viewed as a special case of the more general \textit{three-way MDS} \citep{Carroll1980}, where the three ways correspond to (1) the $n$ data points, (2) the $p^*$ dimensions in the consensus embedding, and (3) the $M$ different distance matrices. 

\method{} thus inherits many desirable diagnostic tools that have been previously developed for INDSCAL and three-way MDS models \citep{deLeeuw2009}. 
For example, one can compute the stress (or loss) per point $i'$, defined as $\sum_{m = 1}^M \sum_{i' < j} \{\Mat{D}_{i'j}^{(m)} - d(\hat{\Mat{W}}^{(m)} \hat{\Mat{Z}_{i'}}, \hat{\Mat{W}}^{(m)} \hat{\Mat{Z}_j})\}^2$, or the stress per candidate dimension reduction method $m'$, defined as $\sum_{i < j} \{\Mat{D}_{ij}^{(m')} - d(\hat{\Mat{W}}^{(m')} \hat{\Mat{Z}_i}, \hat{\Mat{W}}^{(m')} \hat{\Mat{Z}_j})\}^2$. These diagnostics help to identify points or dimension reduction methods that do not fit well with the consensus embedding. 
The optimization problem in CoMDS can also be solved using well-established majorization-minimization (MM) algorithms, as previously detailed in \citet{deLeeuw1980} and \citet{deLeeuw2009} (see Appendix~\ref{app:comds_algorithm} for additional discussion).

More recently, several works have revisited the idea of jointly analyzing multiple distance matrices using MDS-type methods.
In particular, \citet{Chen-JMDS} combines MDS with Wasserstein Procrustes analysis to align and embed data from two different domains into the same joint space. However, this method is only applicable to two views.
\citet{Zhang-UMDS} leveraged an alternative multi-view extension of MDS called Uniform Multidimensional Scaling (UMDS) for clustering while \citet{chen2023similarity} integrated multiple distance matrices for regression tasks.
More closely related to this work is the development of multi-view MDS \citep{Bai-MVMDS}, which is a similar multi-view extension of MDS as in \eqref{eq:comds}, but without the scaling matrices $\Mat{W}^{(1)}, \ldots, \Mat{W}^{(1)}$ and with the addition of view-specific weights. Multi-view MDS, however, was originally developed for general multi-view learning.
For the purposes of identifying a consensus, view-specific weights are typically not needed as the views should ideally be treated equally whereas the scaling matrices $\Mat{W}^{(1)}, \ldots, \Mat{W}^{(1)}$ are necessary to account for possible differences in scale between distance matrices.

\paragraph{Extensions of \method.}
Another key advantage of \method{} is its simplicity and flexibility, which can be easily adapted to a variety of problem settings such as when there are missing data, outliers, and other structural constraints.
In its most general form, \method{} can be viewed as an approach to find the consensus representation $\hat{\Mat{Z}}^*$ via:
\begin{align}
    \hat{\Mat{Z}}^*, &\hat{\Mat{W}}^{(1)}, \ldots, \hat{\Mat{W}}^{(M)} = \nonumber \\
    &\argmin_{\substack{\Mat{Z} \in \R^{n \times p^*}\\ \Mat{W}^{(m)} \in \Omega(p^*)}} \sum_{m = 1}^{M} \sum_{i < j} \omega_{ij}^{(m)} \Big\{ d(\Mat{Z}^{(m)}_i, \Mat{Z}^{(m)}_j) - d(\Mat{W}^{(m)} \Mat{Z}_i, \; \Mat{W}^{(m)} \Mat{Z}_j) \Big\}^2 + \lambda \cdot \text{penalty}, \label{eq:comds_general}
\end{align}
where $\Omega(p^*)$ is some set of allowable $p^* \times p^*$ matrix transformations, $\omega_{ij}^{(m)}$ are weights corresponding to the ($i, j$) sample pair in the $m$-th distance matrix, $d(\cdot, \cdot)$ can be any distance metric, and $\lambda \cdot \text{penalty}$ is some general regularization term.

For example, $\Omega(p^*)$ can be used to allow for more flexible transformations (e.g., rotations \citep{Carroll1972}) beyond just rescaling via diagonal matrices. 
The weights $\omega_{ij}^{(m)}$ can be used to downweight outliers or exclude missing data points (e.g., setting $\omega_{ij}^{(m)} = 0$ if either sample $i$ or $j$ is missing in the $m$-th dimension reduction method). 
Penalties can be added to encourage certain structural properties (e.g., sparsity or smoothness) in the consensus embedding $\Mat{Z}^*$.
Different distance metrics $d(\cdot, \cdot)$ can help to capture different properties of the dimension reduction embeddings---e.g., using a rank-based distance, as in nonmetric MDS \citep{Kruskal-iMDS}, to capture nonlinear structure.
In this work however, we focus on the simplest case of \method, presented in \eqref{eq:comds}, and explore one particularly useful extension of \method{} for dealing with outliers and other large distances in Section~\ref{subsec:locomds}, leaving other extensions to future work.

\subsection{Local Consensus Multidimensional Scaling (\localmethod)}\label{subsec:locomds}

Though \method{} can effectively extract the shared patterns across multiple dimension reduction methods in many cases, there are two important limitations.
First, due to the squared terms in \eqref{eq:comds}, \method{} can be sensitive to outliers in the input dimension reduction methods.
Second, many dimension reduction methods such as t-SNE and UMAP prioritize the preservation of local structure while distorting the overall global structure \citep{Kobak2019}. This leads to meaningless distances between points that are far apart from each other, which can unduly influence \method{}.

To address these limitations, we develop a localized variant of \method{} called Local Consensus Multidimensional Scaling (\localmethod).
While \method{} aims to preserve the pairwise distances between all possible pairs of points, \localmethod{} focuses on preserving only the \textit{local} neighborhood of points (i.e., the small pairwise distances).
More specifically, we build upon the previously-developed Local MDS, proposed in \citet{Chen-LMDS}, and extend it to the multi-view setting to obtain \localmethod, wherein we replace the optimization problem \eqref{eq:comds} in Step 2 above with
\begin{align}
&\hat{\Mat{Z}}^{*}, \hat{\Mat{W}}^{(1)}, \ldots, \hat{\Mat{W}}^{(M)} = \label{eq:locomds} \\ 
&\argmin_{\substack{\Mat{Z} \in \R^{n \times p^*}\\ \Mat{W}^{(m)} \in \text{Diag}(p^*)}} \sum_{m = 1}^{M} \bigg\{ 
    \underbrace{\sum_{(i, j) \in \mathcal{N}^{(m)}_{\pi}} \left( \Mat{D}_{ij}^{(m)} - d(\Mat{W}^{(m)}\Mat{Z}_i, \Mat{W}^{(m)}\Mat{Z}_j) \right)^2}_{\substack{\text{Local Stress}\\\text{(preserve distances between close points)}}}
    - \; t_m \underbrace{\sum_{(i, j) \notin \mathcal{N}^{(m)}_{\pi}} d(\Mat{W}^{(m)}\Mat{Z}_i, \Mat{W}^{(m)}\Mat{Z}_j)}_{\substack{\text{Repulsion Penalty}\\\text{(push apart distant points)}}} \nonumber 
\bigg\}.
\end{align}
Here, $t_1, \ldots, t_M > 0$ are hyperparameters that control the strength of penalty applied to points that are far apart from each other, and $\mathcal{N}^{(m)}_{\pi}$ is the set of neighboring sample pairs in the $m$-th dimension reduction space with distances in the bottom $\pi$-th percentile (i.e., $\{(i, j) \in [n] \times [n] : \Mat{D}_{ij}^{(m)} < Q_{\pi}(\Mat{D}^{(m)})\}$, where $Q_{\pi}(\Mat{D}^{(m)})$ is the $\pi$-th percentile across all entries in $\Mat{D}^{(m)}$). 

At a high level, the objective function in \localmethod{} balances two components: (1) a \textit{local stress} term that encourages the consensus embedding $\hat{\Mat{Z}}^*$ to preserve distances between points that are close to each other in the $M$ dimension reduction embeddings, and (2) a \textit{repulsion penalty} term that encourages points that are far away from each other in the dimension reduction embeddings to remain far away in the consensus embedding, without requiring their exact distances to be preserved. 
This balance is controlled by the hyperparameters $t_1, \ldots, t_M > 0$ (repulsion strength) and $\pi \in (0, 1]$ (neighborhood size). 
Generally, larger values of $t_1, \ldots, t_M$ place greater emphasis on pushing apart distant points relative to preserving local distances while larger values of $\pi$ result in larger neighborhoods and less localization.
We show in Appendix~\ref{app:locomds_algorithm} that the optimization problem in \eqref{eq:locomds} can also be efficiently solved using MM algorithms similar to those used for \method.

\vspace{-0.5em}
\paragraph{Hyperparameter Tuning.} 
In practice, the hyperparameters $t_1, \ldots, t_M > 0$ and $\pi \in (0, 1]$ can be tuned in a data-driven manner similar to \citet{Chen-LMDS}. First, rather than tuning $t_1, \ldots, t_M$ separately, it is often practically easier and still effective to tune a single hyperparameter $\tau$ and to let $t_m$ be a normalized version of $\tau$---namely,
\begin{align*}
    t_m = \frac{\pi}{1 - \pi} \cdot \text{median}_{\mathcal{N}}(\Mat{D}^{(m)}) \cdot \tau,
\end{align*}
where $\text{median}_{\mathcal{N}}(\Mat{D}^{(m)})$ is the median value computed over those pairwise distances in $\Mat{D}^{(m)}$ that lie within the neighborhood $\mathcal{N}_{\pi}^{(m)}$.
Then, given a collection of possible values for $\tau$ and $\pi$, we perform a grid search to select the hyperparameters that best preserve the local neighborhood structure from the original data in the final consensus embedding, as measured by the \textit{adjusted local continuity meta-criterion} (adjusted LCMC). As in \citet{Chen-LMDS}, adjusted LCMC is defined as
\begin{align}
    LCMC_{adj} = \underbrace{\frac{1}{k}\left( \frac{1}{n} \sum_{i=1}^{n} \left| NN_{k}^{X}(i) \cap NN_{k}^{Z}(i) \right| \right)}_{LCMC} - \underbrace{\vphantom{\Bigg|}\frac{k}{n - 1}\vphantom{\Bigg|}}_{\text{adjustment}}, \label{eq:lcmc_adj}
\end{align}
where $k$ is the number of nearest neighbors, $NN_{k}^{X}(i)$ is the set of $k$ nearest neighbors of sample $i$ in the original $\Mat{X}$ space, and $NN_{k}^{Z}(i)$ is the set of $k$ nearest neighbors of sample $i$ in the low-dimensional consensus embedding $\hat{\Mat{Z}}^*$. Intuitively, the LCMC quantifies the proportion of overlapping $k$-nearest neighbors between the original $\Mat{X}$ space and the low-dimensional consensus embedding $\hat{\Mat{Z}}^*$. An adjustment term $\frac{k}{n - 1}$ is then subtracted from the LCMC to account for the expected overlap in $k$-nearest neighbors that would occur by random chance. This adjustment ensures that the LCMC values are comparable across different $k$'s. Higher adjusted LCMC values indicate better preservation of local structure. Thus, $\tau$ and $\pi$ should be chosen to maximize the adjusted LCMC.

It is also important to note that this tuning procedure depends on the choice of neighborhood size $k$ used to compute the adjusted LCMC. If prior knowledge about the data is not available to guide the choice of $k$, we recommend evaluating the adjusted LCMC across a wide range of reasonable $k$ values and selecting the hyperparameters $\tau$ and $\pi$ that yield robust performance across the different $k$ values. For all empircal studies in this work, we selected the $\tau$ and $\pi$ values that maximized the adjusted LCMC across the plurality of $k$ values considered---i.e., $k = 1, 2, 5$, in addition to seven other values, evenly spaced on a logarithmic scale from 10 to either $0.7n$ or the point at which the adjusted LCMC begins to decrease (additional discussion in Appendix~\ref{app:tuning}).

\paragraph{Alternative Neighborhood Definitions.}
We note that there are many possible ways to define the local neighborhoods $\mathcal{N}_{\pi}^{(m)}$ in \localmethod.
For instance, \citet{Chen-LMDS} define the local neighborhoods as the symmetric set of nearest neighbors; that is, $\mathcal{N}_{\pi}^{(m)}$ is the set of pairs $(i, j)$ such that sample $i$ is among the $\pi$-nearest neighbors of sample $j$ or sample $j$ is among the $\pi$-nearest neighbors of sample $i$ in the $m$-th dimension reduction space.
In the consensus dimension reduction context, one concern with this neighborhood definition is that it includes the $\pi$-nearest neighbors for \textit{every} point.
This may include outliers and/or pairs with relatively large distances, which as discussed previously, are often meaningless and not worth preserving in certain dimension reduction methods.
To avoid such issues, we use a distance threshold and choose $\mathcal{N}_{\pi}^{(m)}$ to be the set of points with pairwise distances less than the $\pi$-th percentile of all pairwise distances in $\Mat{D}^{(m)}$, as described above.
This distance-based neighborhood definition not only avoids including pairs of points with nonsensically large distances but also allows for adaptive neighborhood sizes per sample and per dimension reduction method.

\section{Illustrative Simulations}\label{sec:sims}

To help build intuition about how \method{} and \localmethod{} work in practice, we begin by examining their performance in two illustrative simulations, where we know the ground truth.
In Section~\ref{subsec:gaussian}, we examine the performance of \method{} and \localmethod{} in one of the simplest cases, simulating from a mixture of Gaussians.
We then proceed to a more complex nonlinear simulation in Section~\ref{subsec:swiss_roll}, where we simulate data from a Swiss roll manifold and expect improvements in \localmethod{} over \method{} due to the inherent local structure of the underlying manifold.

In both simulations, we compare the performance of \method{} and \localmethod{} to Meta-Spec where we use either UMAP or kPCA as the final dimension reduction method \citep{Ma-Meta}, hereafter referred to as Meta-Spec (UMAP) and Meta-Spec (kPCA), respectively. We choose Meta-Spec (UMAP) and Meta-Spec (kPCA) as these were both used in \citet{Ma-Meta} (and used the same hyperparameters). To also benchmark against existing multi-view methods, we include Multi-SNE, a multi-view extension of t-SNE \citep{Rodosthenous}.
Throughout, we use the same set of 16 dimension reduction methods, as used in \citet{Ma-Meta}, as input into each consensus dimension reduction method.
These 16 inputted dimension reduction methods include principal components analysis (PCA) \citep{Pearson-PCA, Hotelling-PCA}, multidimensional scaling (MDS) \citep{Torgerson-MDS, Torgerson1958}, Kruskal's non-metric MDS (iMDS) \citep{Kruskal-iMDS, Kruskal-iMDS-numerical}, Sammon's mapping (Sammon) \citep{Sammon-Sammon}, locally linear embedding (LLE) \citep{Roweis-LLE}, Hessian LLE (HLLE) \citep{Donoho-HLLE}, isomap \citep{Tenenbaum-isomap}, kernel PCA (kPCA) \citep{Scholkopf-kPCA} with two parameters ($\sigma = $ 0.01 or 0.001), Laplacian eigenmap (LEIM) \citep{Belkin-LEIM}, UMAP \citep{McInnes-UMAP} with two parameters (\textit{number of neighbors} $=$ 30 or 50), t-SNE \citep{vdMaaten-t-SNE} with two parameters (\textit{perplexity} $ = $ 30 or 50), and PHATE \citep{Moon-phate} with two parameters (\textit{number of neighbors} $=$ 30 or 50). 
Further implementation details for each method are provided in Appendix~\ref{app:methods}. 

\subsection{Mixture of Gaussians Simulation}\label{subsec:gaussian}

\begin{figure}
    \centering
    \includegraphics[width=1\linewidth]{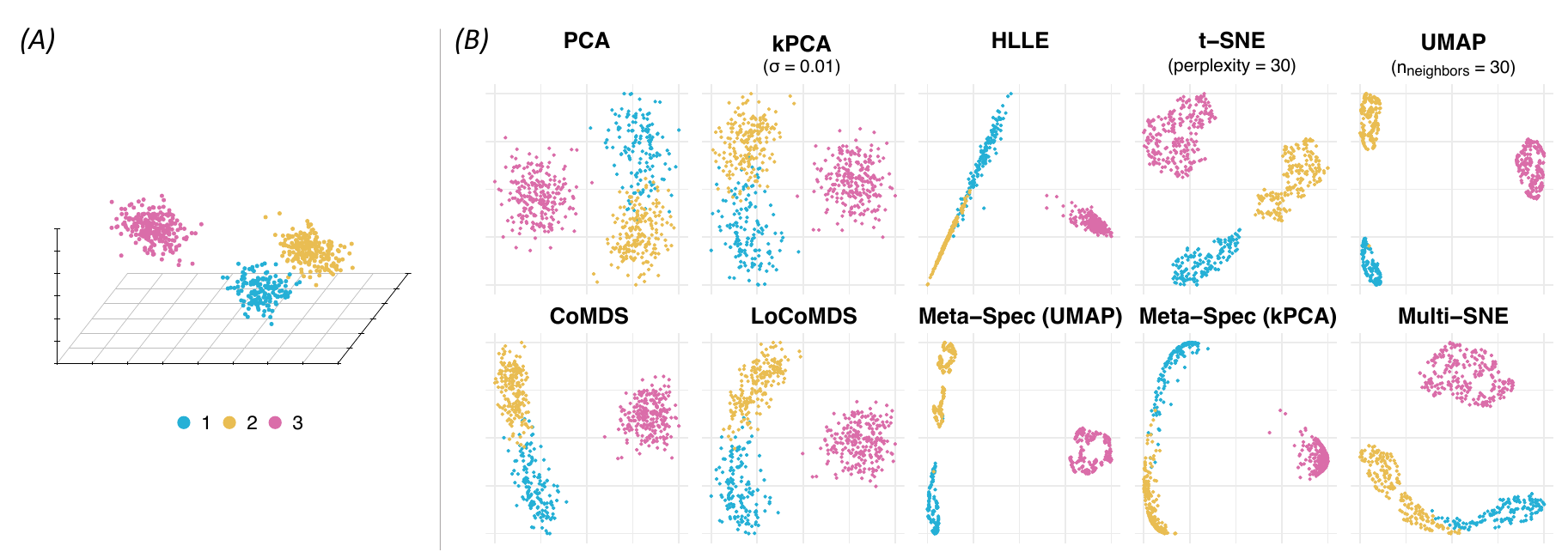}
    \caption{(A) Simulated mixture of Gaussians data. (B) Low-dimensional embeddings obtained from a subset of dimension reduction inputs, \method{}, \localmethod{}, and other existing consensus dimension reduction approaches, applied to the mixture of Gaussians data.}
    \label{fig:gaussian}
\end{figure}

\paragraph{Simulation Setup.} 
In the first simulation, we generated $n = 600$ samples from a mixture of three Gaussians in $p = 3$ dimensions (Figure~\ref{fig:gaussian}A).
There are $150$, $200$, and $250$ samples in clusters 1, 2, and 3, respectively, and samples $\Vec{x}_i$ from each cluster $j$ are generated according to a normal distribution with mean $\Vec{\mu}_j$ and covariance matrix $\Mat{\Sigma} + \sigma^2 \Mat{Z}_j \Mat{Z}_j^{\top}$, where $\Vec{\mu}_1 = (-3, -2, 0)$, $\Vec{\mu}_2 = (2, -4, 1)$, $\Vec{\mu}_3 = (0, 6, 6)$, $\Mat{\Sigma}_{11} = \Mat{\Sigma}_{22} = \Mat{\Sigma}_{33} = 1$, $\Mat{\Sigma}_{12} = 0.3$, $\Mat{\Sigma}_{13} = 0.2$, $\Mat{\Sigma}_{23} = 0.4$, $\sigma = 0.5$, and $\Mat{Z}_j$ is a $3 \times 3$ random standard normal matrix.
Under this setup, clusters 1 and 2 are situated close to each other while cluster 3 is positioned farther away.

\paragraph{Simulation Results.} 
We provide a glimpse into 5 of the 16 input dimension reduction methods in the top row of Figure~\ref{fig:gaussian}B (see Appendix~\ref{app:plots} for full results).
While dimension reduction methods such as PCA, kPCA, and MDS reveal diffuse, Gaussian-shaped clusters, others such as t-SNE and UMAP produce tightly-packed, yet different clusters (e.g., see sole yellow point in UMAP's blue cluster). Still others such as HLLE, PHATE, and LEIM heavily distort the Gaussian shapes altogether.
Such diversity of dimension reduction outputs, present even in this simple simulated dataset, motivates the need to reconcile the different representations.

We thus applied various consensus dimension reduction methods to estimate a consensus visualization from the original 16 dimension reduction methods. 
As shown in Figure~\ref{fig:gaussian}B (bottom row), Meta-Spec (UMAP), Meta-Spec (kPCA), and Multi-SNE all distort the underlying Gaussian shape of the clusters. 
Despite their desired consensus nature, the distorted shapes are idiosyncracies, arising from the underlying choice of dimension reduction method in each of the existing consensus dimension reduction methods (i.e., UMAP in Meta-Spec (UMAP), kPCA in Meta-Spec (kPCA), and t-SNE in Multi-SNE). 
Meta-Spec (UMAP) moreover exacerbates the gap between clusters and reveals additional sub-clustering structure within each large cluster, which is not present in the original data. 
In the case where the true cluster labels were unknown, one might even be misled to believing that there are more than three clusters in the data.
In contrast, both \method{} and \localmethod{} reveal the Gaussian shape of the three clusters, even when many of the input dimension reduction methods introduced their own distortions or spurious, idiosyncratic structures.

\subsection{Swiss Roll Simulation}\label{subsec:swiss_roll}

\begin{figure}
    \centering
    \includegraphics[width=1\linewidth]{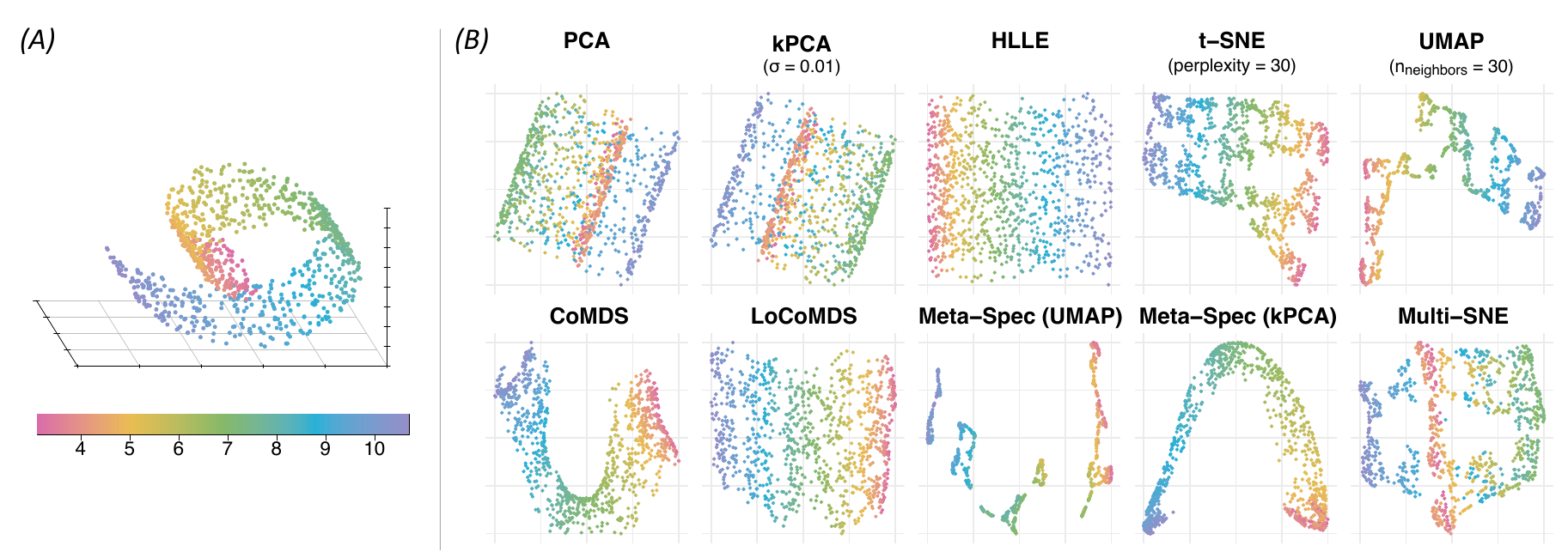}
    \caption{(A) Simulated Swiss roll data. (B) Low-dimensional embeddings obtained from a subset of dimension reduction inputs, \method{}, \localmethod{}, and other existing consensus dimension reduction approaches, applied to the Swiss roll data.}
    \label{fig:swiss_roll}
\end{figure}

\paragraph{Simulation Setup.} 
In the second simulation, we generated $n = 1000$ samples from a Swiss roll manifold in $p = 3$ dimensions (Figure~\ref{fig:swiss_roll}A). The Swiss roll is a popular toy nonlinear transformation, wherein a uniform 2-dimensional flat surface is ``rolled'' up like a Swiss roll pastry in 3-dimensional space. Specifically, points $(x, y, z)$ in the Swiss roll data were generated such that 
\begin{align*}
    x &= t \cos(t) + \varepsilon_x,\\
    y &= t \sin(t) + \varepsilon_y,\\
    z &= u + \varepsilon_z,
\end{align*}
where $t \stackrel{iid}{\sim} \text{Uniform}(\pi, 3.4 \pi)$, $u \stackrel{iid}{\sim} \text{Uniform}(-1, 1)$, and $\varepsilon_x, \varepsilon_y, \varepsilon_z \stackrel{iid}{\sim} N(0, 0.1^2)$. The data was then normalized to have mean zero and unit variance in each $(x, y, z)$ dimension.

\paragraph{Simulation Results.}
As in the mixture of Gaussians simulation, different dimension reduction methods applied to this Swiss roll data output different representations (top row in Figure~\ref{fig:swiss_roll}B; full results in Appendix~\ref{app:plots}).
Unsurprisingly, global linear methods such as PCA fail to capture the underlying nonlinear local structure and instead ``flatten'' the roll onto itself. 
On the other hand, nonlinear neighborhood embedding methods such as t-SNE and UMAP unravel the roll but introduce spurious sub-clusters, which do not exist in the original Swiss roll data where points were simulated uniformly.
In fact, of the 16 dimension reduction methods under consideration, only Isomap, HLLE, and LLE (Figure~\ref{fig:full_swiss}) fully unravel the Swiss roll while also preserving the uniform distribution of points.

Turning now to the consensus visualizations in Figure~\ref{fig:swiss_roll}B (bottom row), Meta-Spec (UMAP) inherits and even exacerbates spurious sub-clusters seen in the original UMAP while Multi-SNE fails to unravel the Swiss roll. 
Meta-Spec (kPCA) performs well but still introduces distortions, as the uniform spacing and width of the roll are not fully captured. 
This uniform spacing and equal width are slightly better preserved by \method{}, which unravels the Swiss roll structure but as a U-shape. 
Such U-shape arises due to the inclusion of all pairwise distances in \method{}; as a result, \method{} pulls together the two ends of the Swiss roll (forming the U-shape) in an attempt to preserve the close proximity of the reds and the purples (i.e., the two ends of the Swiss roll) observed in dimension reduction methods such as PCA and kPCA. 
In contrast, by focusing on the local neighborhood structure, \localmethod{} is able to unfold the Swiss roll more effectively, revealing its intrinsic 2D structure even when the large majority of input dimension reduction methods fail to do so.
Moreover, this highlights the adaptability and advantage of \localmethod{} over \method{} when the underlying data structure is inherently local, as in the Swiss roll manifold.

\section{Real-Data Applications}\label{sec:case_studies}
To further demonstrate the effectiveness of \method{} and \localmethod{}, we next examine their performance alongside existing approaches across nine real-world datasets, summarized in Table~\ref{tab:dataset_info} with more detailed descriptions in Appendix~\ref{app:datasets}. These datasets span a wide range of domains, including agriculture, biology, and astrophysics. For each dataset, we again apply the same set of 16 dimension reduction methods as in Section~\ref{sec:sims} to obtain a diverse set of low-dimensional visualizations and subsequently apply \method{}, \localmethod{}, Meta-Spec (UMAP), Meta-Spec (kPCA), and Multi-SNE to produce a consensus visualization. Hyperparameters and implementation details are the same as in Section~\ref{sec:sims} (details in Appendix~\ref{app:methods}).

In Section~\ref{subsec:olive}, we begin by closely examining one dataset---namely, the olive oil dataset---in order to visually illustrate and highlight key intuitions behind our proposed methods and their advantages in practice (as opposed to simulated settings shown previously in Section~\ref{sec:sims}). 
In Section~\ref{subsec:eval}, we then proceed to more systematically evaluate and quantify the performance of the consensus dimension reduction methods across all nine datasets.

\begin{table}[h!]
\centering
\renewcommand{\arraystretch}{1.1}
\small
\rowcolors{2}{white}{lightgray}
\begin{tabularx}{\textwidth}{c c c >{\raggedright\arraybackslash}X >{\centering\arraybackslash}p{2.5cm}}
\toprule
\textbf{Dataset} & \textbf{$n$} & \textbf{$p$} & \centering \textbf{Description} & \textbf{Clusters} \\
\midrule
HIV & 1400 & 70 & 
scRNA-seq data from peripheral blood mononuclear cells from HIV-infected individuals \citep{Kazer}&
7 cell types \\
4EQ & 1200 & 100 &
scRNA-seq data from four human peripheral blood mononuclear cell subtypes \citep{Duo}&
4 cell types \\
8EQ & 1600 & 50 &
scRNA-seq data from eight human peripheral blood mononuclear cell subtypes \citep{Duo} &
8 cell types \\
Trajectory & 421 & 500 &
scRNA-seq data of embryonic stem cells differentiating into primitive endoderm cells \citep{Hayashi-trajectory} &
5 periods after differentiation \\
Cycle & 288 & 1147 &
scRNA-seq data from mouse embryonic stem cells \citep{Buettner-cycle} &
3 cell-division stages \\
Olive Oil & 572 & 8 &
Chemical measurements of olive oil samples \citep{Forina} &
9 production regions \\
Wheat & 210 & 7 &
Measurements of geometric properties of wheat kernels \citep{Charytanowicz-wheat}&
3 wheat varieties \\
Wholesale & 440 & 6 &
Annual spending on various product types by wholesale customers \citep{UCIWholesale}& 
2 client types \\
Star & 1000 & 6 &
Photometric data of different stellar objects from Sloan Digital Sky Survey \citep{data-star} &
3 astronomical object types \\
\bottomrule
\end{tabularx}
\caption{Summary of real-world datasets, including the number of observations ($n$), number of features ($p$), a brief description, and the number of clusters in each dataset.}
\label{tab:dataset_info}
\end{table}

\subsection{Olive Oil Data Case Study}
\label{subsec:olive}

In Figure~\ref{fig:olive}, we present a subset of the 16 input dimension reduction methods (full results in Appendix~\ref{app:plots}) and the visualizations obtained from various consensus dimension reduction methods when applied to the olive oil dataset, containing chemical measurements of 572 olive oil samples from nine different regions in Italy.

As in the illustrative simulations in Section~\ref{sec:sims}, we observe a wide spectrum of representations produced by the different dimension reduction methods (Figure~\ref{fig:olive}, top row), which can be difficult to reconcile and interpret in practice.
In particular, while PCA and kPCA produce diffuse clusters, t-SNE and UMAP yield tight, well-separated clusters that are also different from each other. 
Meanwhile, HLLE (which performed well on the Swiss roll) is clearly dominated by two extreme outliers that are not observed in other dimension reduction methods. 
To further complicate matters, the global positioning of the clusters differs across dimension reduction methods, particularly when comparing the position of the Sardinia oils (yellow and green) relative to the other oils.

This global structure remains a major difference between consensus dimension reduction methods (Figure~\ref{fig:olive}, bottom row). While Meta-Spec (kPCA) and Multi-SNE place Sardinia close to West Liguria (maroon) but far from South Apulia (orange), \method{} and \localmethod{} preserve the global structure seen in methods such as PCA, placing the Sardinia oils (yellow and green) between the South Apulia (orange) and West Liguria (maroon). 
Notably, Sardinia and South Apulia share similar subtropical climates and experience similar levels of annual precipitation, air temperatures, potential evapotranspiration, and soil aridity \citep{Costantini2013}.
Thus, the global structure revealed by \method{} and \localmethod{} appears to be more consistent with domain knowledge, compared to the structure shown in the Meta-Spec (UMAP), Meta-Spec (kPCA), and Multi-SNE. 
Finally, we highlight that the main difference between \method{} and \localmethod{} lies in their treatment of outliers: while the two outliers from HLLE remain in \method{}, \localmethod{} is more robust to these outliers and does not preserve them given that they are not present in any of the other input dimension reduction methods.

\begin{figure}
    \centering
    \includegraphics[width=1.0\linewidth]{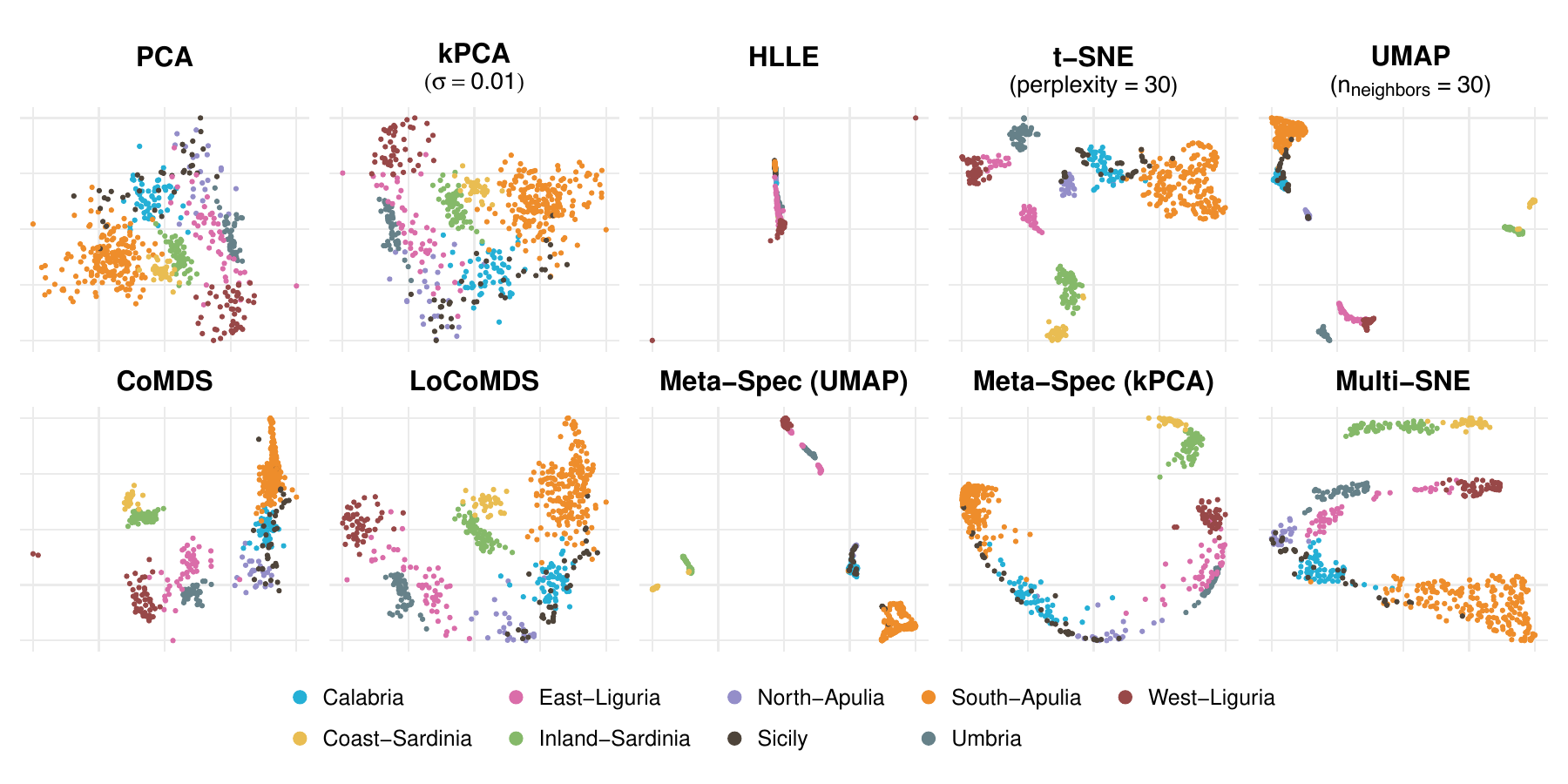}
    \caption{
        Low-dimensional embeddings obtained from a subset of dimension reduction inputs, \method{}, \localmethod{}, and other existing consensus dimension reduction approaches, applied to the olive oil dataset.
    }
    \label{fig:olive}
\end{figure}

Though we have focused on the olive oil dataset here, similar patterns can also be observed across the other eight datasets and are thus deferred to Appendix~\ref{app:plots}.

\subsection{Quantifying Performance of Consensus Dimension Reduction Methods}\label{subsec:eval}

Moving beyond visual inspection, we also seek to conduct a more systematic evaluation of performance across the different datasets according to quantifiable metrics. 
Here, since the true underlying structure of these real datasets is typically unknown, quantifying the performance of dimension reduction methods is admittedly challenging with no universally agreed-upon evaluation framework. 

Recent works nevertheless have made progress towards developing a comprehensive evaluation framework for dimension reduction \citep{Espadoto, Huang-evaluation}. We leverage these works and quantify the performance of the consensus dimension reduction methods under study according to three main criteria: 
\begin{enumerate}
    \item \textit{Global and Local Structure Preservation}: measured by how well the low-dimensional representations preserve both the global and local structure of the original (higher-dimensional)~data;
    \item \textit{Supervised Evaluation and Prediction Accuracy}: measured by how well the low-dimensional representations preserve known substantively-meaningful clusters in the original data; and
    \item \textit{Stability}: measured by how stable the consensus visualizations are to the choice of dimension reduction methods and their hyperparameters.
\end{enumerate}
In what follows, we demonstrate that \method{} and \localmethod{} (1) strike an effective balance between preserving both global and local structure in the data, (2) maintain high predictive accuracy and good separation of known relevant clusters, and (3) yield stable visualizations that are not only robust to the choice of input dimension reduction methods but can also be used to reveal and mitigate hyperparameter instability in existing dimension reduction methods.

\subsubsection{Global and Local Structure Preservation}\label{subsec:eval_structure}

As advocated by \citet{Huang-evaluation}, a good dimension reduction method should preserve both global and local structures that are present in the original data $\Mat{X}$.
In other words, points that are close (``local'') in the original space should remain close in the low-dimensional space while points that are far apart (``global'') in the original space should also remain far apart in the low-dimensional space. Following \citet{Huang-evaluation}, we evaluate the global and local structure preservation for each consensus dimension reduction method under study using complementary metrics detailed next.

\paragraph{Global Structure Preservation Evaluation Metrics.}
To measure how well the global structure is preserved in a dimension reduction plot, we evaluate (i) the \textit{random triplet accuracy}, defined as the proportion of randomly sampled triplets (i.e., sets of 3 points), for which the relative ordering of pairwise distances within each triplet is preserved between the original space and the low-dimensional space, and (ii) the \textit{Spearman correlation} between the pairwise distances in the original space and their corresponding pairwise distances in the dimension-reduced space.
Higher random triplet accuracy and Spearman correlation values indicate that the embedded points better preserve the global structure of the original data.

\paragraph{Local Structure Preservation Evaluation Metrics.}
To evaluate local structure preservation, we use the (unadjusted) local continuity meta-criterion (LCMC) score, previously defined in \eqref{eq:lcmc_adj}, for small values of $k$ (ranging between 2 and 20 with a step size of 3). 
The LCMC score (sometimes referred to as neighborhood retention) measures the proportion of overlap between the $k$-nearest neighbors of each point in the original space and in the dimension-reduced space, averaged across all points. Higher LCMC values indicate better preservation of local structure.

\paragraph{Global and Local Structure Preservation Results.}

\begin{figure}
    \centering
    \includegraphics[width=1\linewidth]{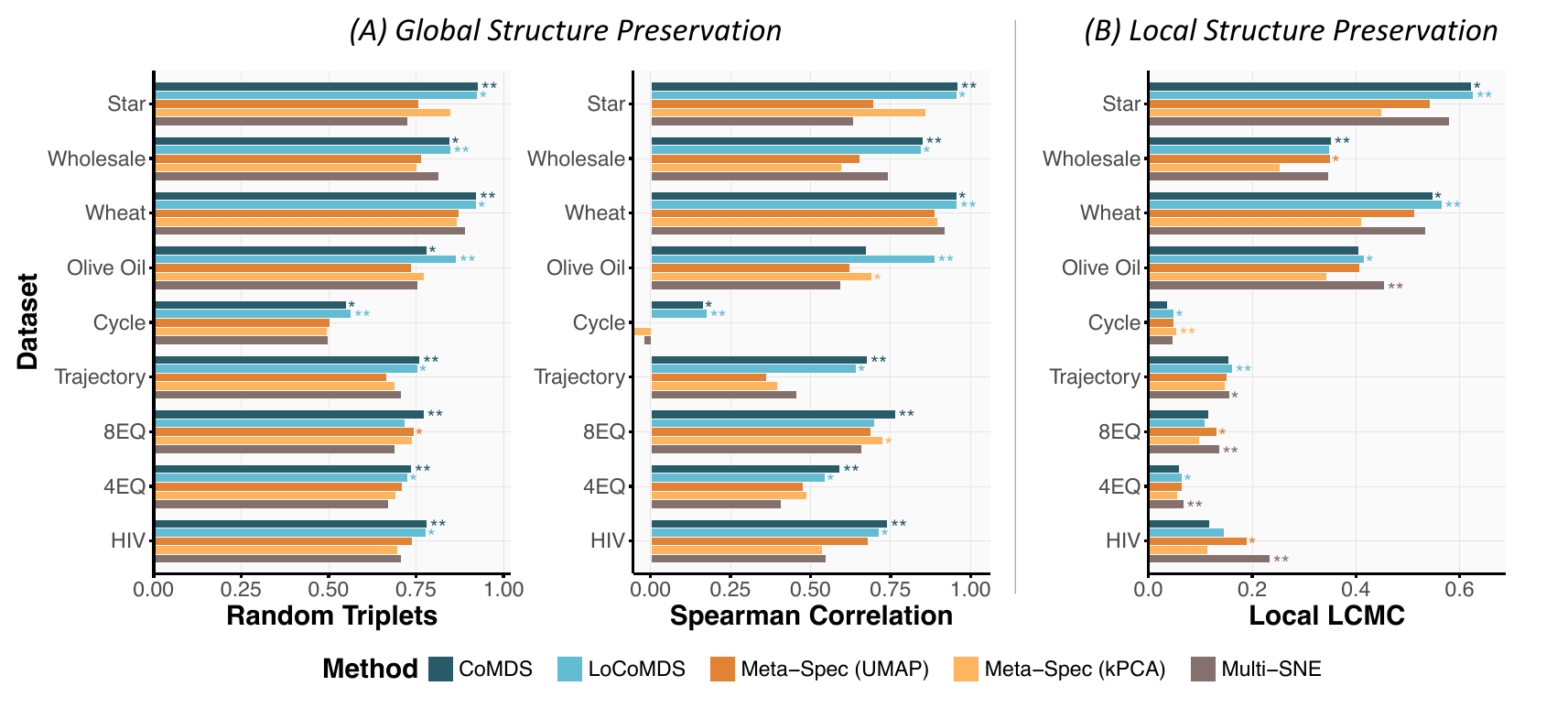}
    \caption{
        (A) Global and (B) local structure preservation results across all nine datasets. Higher values indicate better preservation of the respective structure. For each dataset and metric, the method with the best performance is denoted by two asterisks while the second-best method is denoted by one asterisk.
    }
    \label{fig:ALLEval}
\end{figure}

We summarize the global and local structure preservation results across all nine datasets in Figure~\ref{fig:ALLEval}. Here, we report the local LCMC, averaged across all small $k$ values considered, and defer the full LCMC curves to Appendix~\ref{app:eval_structure}.

With respect to global structure preservation, \method{} and \localmethod{} are consistently the top-two performing methods across almost all datasets and metrics (random triplet accuracy and Spearman correlation). Additionally, even though \localmethod{} is more tailored towards preserving the local structure of the data, \localmethod{} sometimes outperforms \method{} in preserving the global structure.
As alluded to in Sections~\ref{sec:sims} and \ref{subsec:olive}, this typically occurs when there are outliers, introduced by methods such as HLLE, which favor \localmethod{}'s robustness against such points.

Meanwhile, Meta-Spec (UMAP) and Multi-SNE primarily focus on preserving the local structure of the data, similar to UMAP and t-SNE themselves. 
Thus unsurprisingly, Meta-Spec (UMAP) and Multi-SNE yield the highest local structure preservation scores on several datasets, but at the cost of poor global structure preservation, exhibiting the worst random triplet accuracy and Spearman correlation scores.
Still, \method{} and \localmethod{} remain highly competitive in their local structure preservation, with \localmethod{} generally yielding slightly better local structure preservation than \method{}.
In fact, \method{} and \localmethod{} yield the best two local LCMC scores on the Wheat and Star datasets.
If we average each method's local LCMC ranking across all datasets (Table~\ref{tab:method_ranks_structure}), we further find that \localmethod{} achieves the best average rank (even better than Meta-Spec (UMAP) and Multi-SNE), highlighting its ability to strike a balance between preserving both global and local structure and more importantly, achieve strong performance across both criteria.

\subsubsection{Supervised Evaluation and Prediction Accuracy}\label{subsec:eval_prediction}

In addition to preserving the global and local data structure, we expect a good dimension reduction method to preserve known substantively-meaningful patterns from the original data.
Here, due to the lack of ground truth about the underlying low-dimensional structure of the data, we use known cluster labels, associated with each dataset (Table~\ref{tab:dataset_info}), as a proxy for these substantively-meaningful patterns and evaluate how well we can predict these cluster labels based upon the low-dimensional representations.
Note these cluster labels were not used in the dimension reduction process itself and that each dataset was chosen because it contains known cluster labels.

\paragraph{Prediction Evaluation Metrics.}
More specifically, we first randomly split the consensus dimension reduction embeddings data into training (80\%) and test (20\%) sets. 
Using the training data, we fit a random forest (RF) \citep{Breiman-rf} with 300 trees (alongside other default hyperparameters) to predict the associated cluster labels using the low-dimensional consensus embeddings. 
RF was chosen as it is both nonparametric and requires minimal hyperparameter tuning.
We then evaluated the prediction performance according to the (multi-class) area under the receiver operating characteristic curve (AUROC) on the test set.
We repeated the random train-test split 100 times and report the average AUROC across these 100 replicates.

\paragraph{Prediction Results.} 

We summarize the RF prediction accuracies across all nine datasets in Table~\ref{tab:method_ranks_rf}A, showing the number of datasets for which each method performed the best (i.e., rank 1), second-best (i.e., rank 2), etc. 
While generally different consensus dimension reduction methods yielded better prediction accuracy for different datasets, Multi-SNE had the best overall average ranking across all datasets, followed by \localmethod{} in second.
However, as shown in Figure~\ref{fig:rf_eval}A, the differences in AUROC between consensus dimension reduction methods are often very small, and the overall AUROCs are all close to 1. This indicates that generally all consensus dimension reduction methods under study are preserving substantively-meaningful patterns in the data well.

More interesting though is how the RF behaves on samples that are misclassified or assigned to a wrong cluster. 
As the visualizations in previous sections may suggest, Multi-SNE, Meta-Spec (UMAP), and Meta-Spec (kPCA) tend to produce very tight, well-separated clusters, whether or not such clusters truly exist in the data.
This can lead to overly-confident, but wrong predictions.
To more rigorously quantify this notion, for each dataset, we examined the samples that were misclassified by the RF in all five consensus dimension reduction methods and averaged their predicted probabilities of their assigned (but wrong) class.
High predicted probabilities indicate that the RF was highly confident in its wrong prediction. 
As seen in Table~\ref{tab:method_ranks_rf}B (and Figure~\ref{fig:rf_eval}B), Meta-Spec (UMAP) was almost always the most confident in its wrong predictions, followed by Meta-Spec (kPCA) and Multi-SNE.
Such overconfidence is problematic in practice, as it may mislead practitioners into highly trusting the cluster assignments, which are visually obvious in the low-dimensional embeddings, but are actually incorrect.
In contrast, \method{} and \localmethod{} on average yielded the least confident predictions for these misclassified samples, suggesting that these methods better reflect the ambiguity and uncertainty in the misclassified points while still achieving competitively high prediction accuracy.

\begin{table}[h!]
\centering
\renewcommand{\arraystretch}{1.2}
\small
\begin{tabular}{l c c c c c c}
\toprule
\multicolumn{7}{c}{\textit{(A)} \textbf{RF AUROC}} \\
\textbf{Method} & Rank 1 & Rank 2 & Rank 3 & Rank 4 & Rank 5 & \textbf{Average Rank}\\
\midrule
\method & 1 & 1 & 5 & 1 & 1 & 3.00 \\
\localmethod & 0 & 5 & 2 & 1 & 1 &2.78 \\
\metau & 3 & 1 & 1 & 2 & 2 & 2.89 \\
\metak & 0 & 0 & 1 & 3 & 5 & 4.44 \\
\msne & 5 & 2 & 0 & 2 & 0 & \textbf{1.89} \\
\bottomrule
\end{tabular}

\vspace{0.3cm}

\begin{tabular}{l c c c c c c}
\toprule
\multicolumn{7}{c}{\textit{(B)} \textbf{RF Predicted Probability
of Wrong Class}} \\
\textbf{Method} & Rank 1 & Rank 2 & Rank 3 & Rank 4 & Rank 5 & \textbf{Average Rank}\\
\midrule
\method & 3 & 2 & 3 & 1 & 0 & 2.22 \\
\localmethod & 3 & 4 & 1 & 1 & 0 & \textbf{2.00} \\
\metau & 0 & 0 & 1 & 2 & 6 & 4.56 \\
\metak & 0 & 2 & 2 & 4 & 1 & 3.44 \\
\msne & 3 & 1 & 3 & 1 & 2 & 2.78 \\
\bottomrule
\end{tabular}

\caption{
    Number of datasets (out of 9) for which each method performed the best (i.e., rank 1), second-best (i.e., rank 2), etc, according to (A) the random forest (RF) AUROC and (B) the average RF predicted probability of the wrong class for misclassified samples. Rank 1 corresponds to the best-performing method for a given dataset while Rank 5 corresponds to the worst-performing method.
}
\label{tab:method_ranks_rf}
\end{table}

\subsubsection{Stability Case Studies}\label{subsec:eval_stability}

Finally, as discussed in previous works \citep{Huang-evaluation,yu2020veridical}, a good dimension reduction method should be reasonably stable with respect to different modeling choices (e.g., hyperparameters).
This is particularly important in practice given that validating findings from dimension reduction methods is frequently difficult, if not impossible.
We thus examine how the visualizations from consensus dimension reduction methods change with respect to their main modeling choice, that is, the choice of dimension reduction methods to use as input.
We then show how our consensus dimension reduction framework can further be used to mitigate the well-known instability of popular dimension reduction methods such as UMAP with respect to their main modeling choice, that is, the choice of hyperparameters.

\paragraph{Stability to Method Choice.}
To understand how the consensus visualizations change with respect to the choice of input dimension reduction methods, we consider the hypothetical scenario in which a practitioner is trying to decide between two sets of input dimension reduction methods: (i) a set of two ``base'' dimension reduction methods and (ii) a set of the same two base methods plus one additional method that---unknowing to the practitioner---provides a relatively ``poor'' representation of the data.
For the base methods, we use the two dimension reduction methods that yield the highest eigenscores \citep{Ma-Meta} among the 16 candidate methods considered in this paper (see Appendix~\ref{app:stability} for other choices of base methods). Then in an attempt to introduce the most contamination or instability between the two sets of candidates, we use the method that yields the lowest eigenscore as the additional method.
For all datasets considered in this work, HLLE happened to be the method with the lowest eigenscore, which we have seen to produce skewed dimension reduction embeddings with severe outliers (see Appendix~\ref{app:plots}).

\begin{figure}
    \centering
    \includegraphics[width=0.9\linewidth]{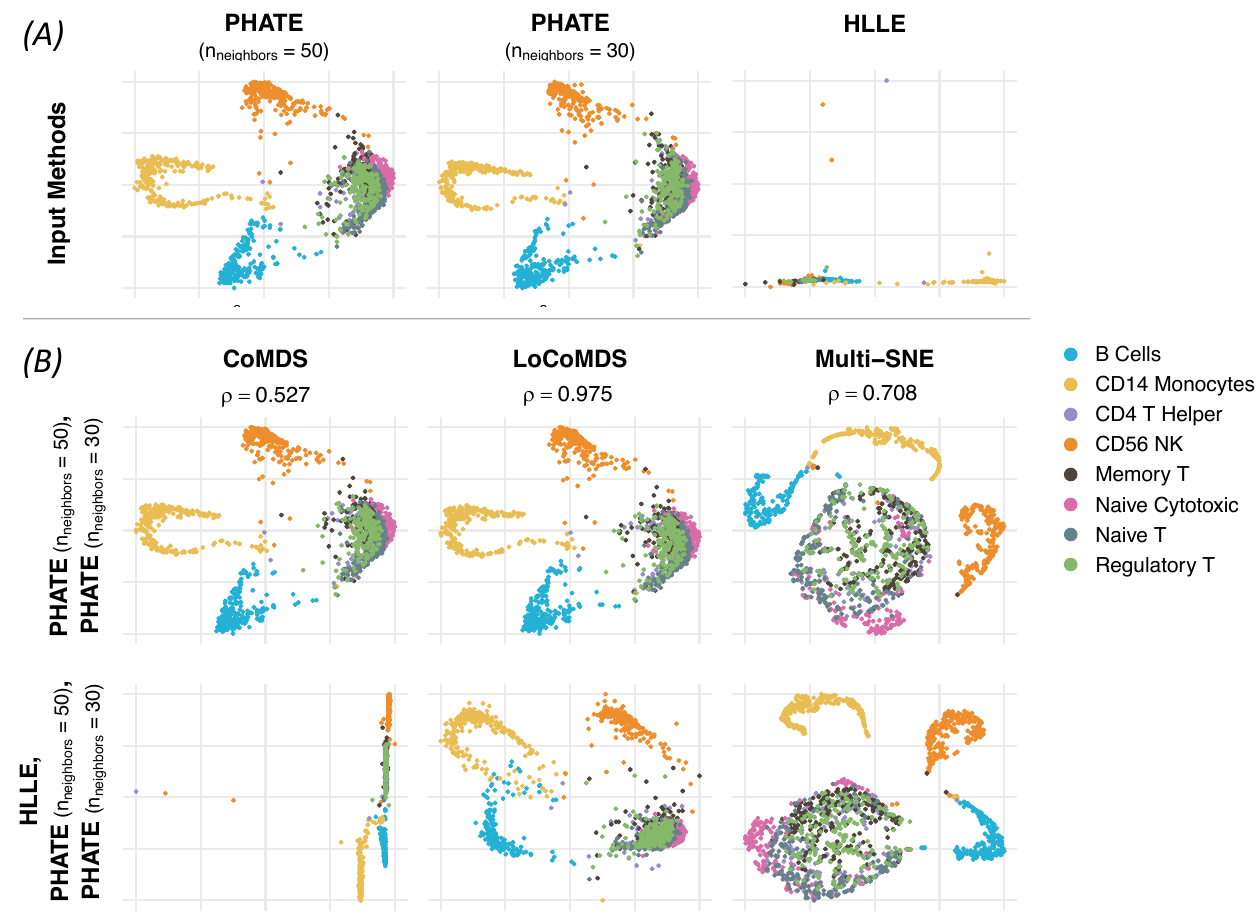}
    \caption{
        (A) Low-dimensional embeddings from PHATE embeddings with a number of neighbors 30 and 50 as well as HLLE on the 8EQ dataset. 
        (B) Low-dimensional embeddings obtained from \method{}, \localmethod{}, and \msne{} on the 8EQ dataset using (top) the two PHATE embeddings as input, compared to (bottom) the same two PHATE embeddings plus HLLE as input. The Mantel test statistics, measuring the correlation between the two embeddings for each method, are also reported and denoted by $\rho$.}
    \label{fig:stability_method_8eq}
\end{figure}

Focusing on the 8EQ dataset here as a case study, Figure~\ref{fig:stability_method_8eq} shows the consensus visualizations produced by \method{}, \localmethod{}, and Multi-SNE, applied to the two different sets of input dimension reduction methods. 
We highlight two key observations.
First, the two base methods (i.e., PHATE with a number of neighbors of 30 and 50 for this dataset) are almost identical to each other as seen in Figure~\ref{fig:stability_method_8eq}A. Consequently, when applied to only these base methods, \method{} and \localmethod{} preserve the shared structure and output similar consensus visualizations to the inputs, as expected (Figure~\ref{fig:stability_method_8eq}B, top row). In contrast, Multi-SNE introduces its own idiosyncratic structure, yielding a visually-different consensus output from the inputs despite their nearly-identical structure.

Second, when HLLE is added (Figure~\ref{fig:stability_method_8eq}B, bottom row), the global structure in Multi-SNE---particularly, the positions of the B cells (blue), CD14 monocytes (yellow), and CD56 NK cells (orange)---changes substantially compared to when only using the two base methods as input.
\method{} is also heavily influenced by outliers in HLLE and is generally not recommended when such outliers are present.
\localmethod{}, on the other hand, appears to be the most robust to the inclusion of HLLE and yields similar visualizations regardless of the choice of input methods.
This is to say that the unwanted idiosyncratic structure introduced by HLLE minimally affects the consensus visualization from \localmethod{}, which continues to largely reflect the shared structure among the two base methods. 

We can further quantify this similarity between results with and without HLLE via the Mantel test statistic, which measures the correlation between two distance (or dissimilarity) matrices \citep{Mantel}.
The Mantel test statistics for Multi-SNE, \method{}, and \localmethod{} are 0.708, 0.527, and 0.975, respectively, confirming that \localmethod{} is the most stable to the choice of input dimension reduction methods. 
In settings where a practitioner is unsure about which dimension reduction methods to use as input and may unknowingly include a poor dimension reduction method, this stability property is highly desirable and reduces the burden of method selection by the practitioner.
Results using other datasets similarly demonstrate this stability of \localmethod{} and are deferred to Appendix~\ref{app:stability}.

\paragraph{Stability to Hyperparameter Choice.}
Another important practical advantage of our proposed consensus dimension reduction framework is the ability to mitigate the sensitivity of existing dimension reduction methods to the choice of hyperparameters.
For instance, UMAP is well-known to be sensitive to the number of neighbors hyperparameter, which controls the size of the local neighborhood that UMAP uses to approximate the manifold structure of the data. 
Fortunately, our consensus dimension reduction framework can be naturally used to help address this form of instability.

\begin{figure}
    \centering
    \includegraphics[width=0.9\linewidth]{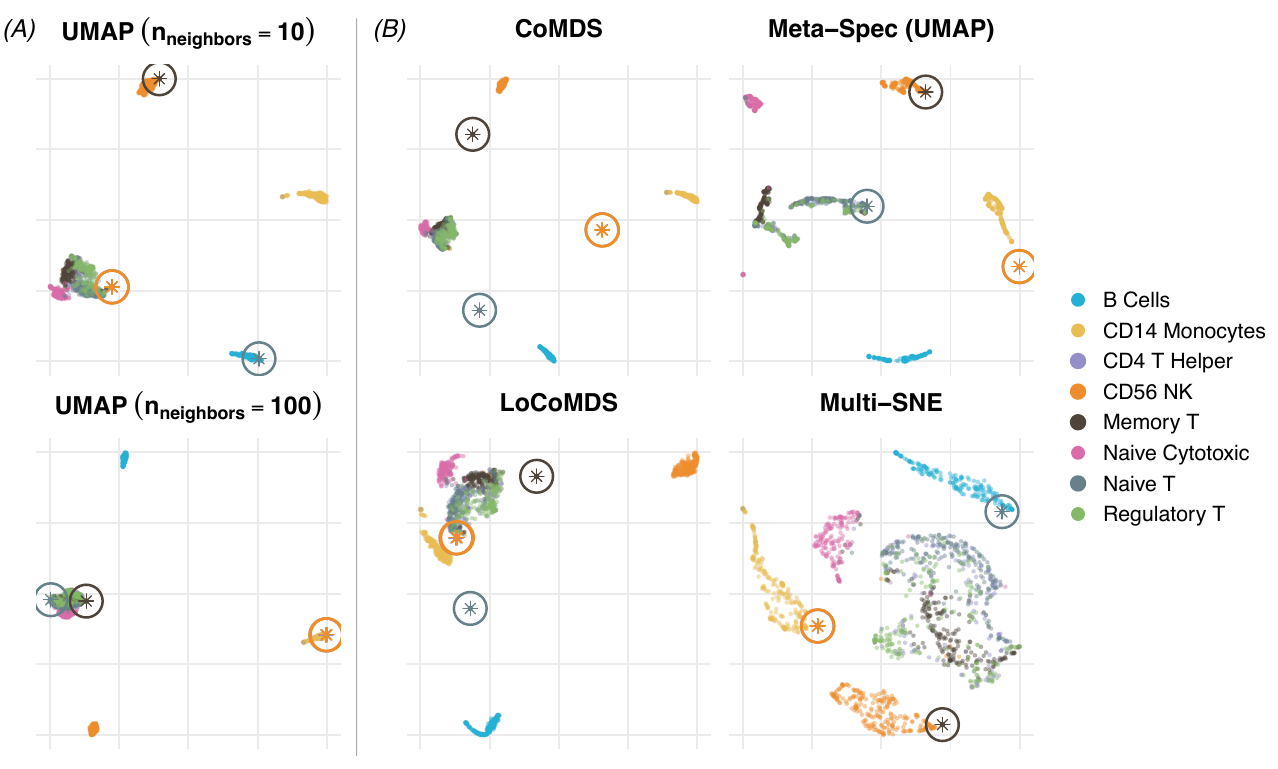}
    \caption{(A) Low dimensional visualizations obtained from UMAP with the number of neighbors of 10 and 100 on 8EQ data. (B) Low-dimensional visualizations obtained from consensus dimension reduction methods using these two UMAP representations as input. Points circled and marked with asterisks are those that are assigned to noticeably different clusters under the two UMAP hyperparameter settings.}
    \label{fig:stability_param_8eq}
\end{figure}

To illustrate this, Figure~\ref{fig:stability_param_8eq} shows the UMAP visualizations of the 8EQ data under two different choices of the number of neighbors hyperparameter (10 and 100). 
Notably, there are several points (circled and marked as asterisks) that are clearly assigned to different clusters under the two hyperparameter settings. 
For example, a memory T cell (brown asterisk) is firmly placed within the CD56 NK cluster when using 10 nearest neighbors but is in the large T cell cluster when using 100 nearest neighbors. 
If these cells are unlabeled, practitioners would be unsure about which cluster these cells truly belong to.

In theory, consensus dimension reduction methods can help address this issue by combining the information from both UMAP visualizations to produce a consensus representation as shown in Figure~\ref{fig:stability_param_8eq}B. 
However, existing methods such as Meta-Spec (UMAP) and Multi-SNE force these ambiguous points into one cluster or the other, failing to reflect the uncertainty about these points.

\method{}, on the other hand, illuminates the ambiguity of these points by placing them near the midpoint of the two clusters.
This behavior and information may even help practitioners identify samples that require further investigation.
\localmethod{} similarly places these ambiguous points between the two clusters, but with one apparent difference: the local nature of \localmethod{} allows itself to rearrange the global cluster structure, pulling the CD14 monocyte cluster closer to the large T cell cluster even though this was not seen in either UMAP visualization. 
Here, since one CD56 NK cell (orange asterisk) appears in the T cell cluster when using 10 nearest neighbors but in the CD14 monocyte cluster using 100 nearest neighbors, \localmethod{} uses this information to pull these two clusters closer together and places this ambiguous point between them.
Lastly, we highlight that \method{} and \localmethod{} again avoid confidently misclassified points. Each cluster is well separated with no points being placed deep within the wrong cluster, illustrating the practical utility of focusing on the consensus or shared structure across different hyperparameter settings.

\section{Discussion}\label{sec:discussion}
Regardless of whether the data was simulated from the simplest of cases or taken from some noisy, complex real-world application, \textit{every} empirical example presented in this work exhibited a wide spectrum of low-dimensional representations, depending on the dimension reduction method. 
Such diversity in results is initially troubling for practitioners who aim to use dimension reduction methods to further their scientific understanding of a dataset and generate trustworthy hypotheses for future study.
However, this diversity also presents an important opportunity: in line with the stability principle and the driving philosophy behind multi-view learning, findings consistently observed across multiple reasonable analysis choices (e.g., dimension reduction methods) are more likely to be reliable and trustworthy than those observed from only a single view.
We hence built upon this idea and developed Consensus Multidimensional Scaling (\method{}) alongside its local variant, Local Consensus Multidimensional Scaling (\localmethod{}), to enhance the reliability of dimension reduction analyses. Our proposed methods yield a more robust low-dimensional representation by focusing on the shared, or consensus, structure across multiple dimension reduction ``views'' of the same data. To facilitate the adoption of our proposed methods, we provide an open-source R package on GitHub at \url{https://github.com/tiffanymtang/coMDS}.

Furthermore, given the widespread use of dimension reduction methods across applications and scientific domains, we emphasize that this work is only the starting point for improving the reliability of dimension reduction analyses.
A particularly important avenue for future work includes the improved evaluation of dimension reduction methods.
Such evaluation, while inherently challenging due to the unsupervised nature of dimension reduction, is crucial not only for assessing the performance of our proposed methods but more importantly, for guiding the selection of the input dimension reduction methods into \method{} and \localmethod{}.
Though the broad array of 16 dimension reduction methods used in this work provides an effective baseline, yielding strong results here, our proposed consensus dimension reduction methods are designed to learn the shared structure among the input methods.
Thus, if the input dimension reduction representations are of poor quality, the resulting consensus representations from \method{} and \localmethod{}, which extract what is shared from these input dimension reduction methods, will also inevitably be of poor quality.

We lastly highlight the flexibility of our consensus dimension reduction framework as it opens up exciting avenues for future research.
In particular, our framework can be extended to include general regularization terms, weights, or constraints to induce desired properties in the consensus representation, such as sparsity or smoothness.
Missing data is another common challenge in real-world datasets and can be easily handled in our framework using a missingness indicator as sample-specific weights.
Moreover, while our current focus has been on tabular data, the proposed framework is not limited to such data types. Our framework can be used to combine any number and/or types of embedding techniques (e.g., autoencoders for images, large language model embeddings for text, or learned graph representations for network data), making it a flexible and powerful foundation for more reliable visualizations and discoveries in scientific applications and beyond.

\bibliographystyle{chicago}
\bibliography{bibliography}
\clearpage

\appendix
\renewcommand{\thefigure}{A.\arabic{figure}} 
\setcounter{figure}{0} 

\renewcommand{\thetable}{A.\arabic{table}}
\setcounter{table}{0}

\renewcommand{\theequation}{A.\arabic{equation}}
\setcounter{equation}{0}

\begin{center} 
\Large
\textbf{Appendix: ``Consensus dimension reduction via data integration''}
\end{center} 

\section{Optimization Algorithms for \method{} and \localmethod{}}\label{app:algorithms}

To solve the optimization problems (2) and (4) (from the main text) in \method{} and \localmethod{} respectively, we leverage majorization-minimization (MM) algorithms, which have been widely used to optimize MDS-based problems and their constrained variants \citep{deLeeuw1980}.
For details and a comprehensive discussion of these MM algorithms for MDS-based methods, we refer interested readers to \citet{Commandeur} and \citet{deLeeuw2009}.

Below, we outline an MM algorithm to solve (2) (from the main text) in \method{} in Section~\ref{app:comds_algorithm}. This algorithm has been previously introduced and detailed in \citet{Commandeur} and \citet{deLeeuw2009}. Hence, we only provide a brief overview of the main steps here to keep this work self-contained and to show how this algorithm can be extended to solve (4) (from the main text) in \localmethod{} in Section~\ref{app:locomds_algorithm}.

\subsection{\method{} Optimization Algorithm}\label{app:comds_algorithm}

To solve the optimization problem (2) (from the main text) in \method{}, first note that it can be equivalently viewed as an ordinary multidimensional scaling problem with a constraint. More formally, let $\Mat{D}$ be the $Mn \times Mn$ block diagonal matrix with the individual distance matrices $\Mat{D}^{(1)}, \ldots, \Mat{D}^{(M)}$ on the diagonal, and let $\hat{\Mat{Z}}^*_{MDS}$ be the solution to the following constrained MDS problem:
\begin{align}
    \hat{\Mat{Z}}^*_{MDS} = \argmin_{\tilde{\Mat{Z}} \in \R^{Mn \times p^*}} \sum_{i < j} \Big\{ \Mat{D}_{ij} - d(\tilde{\Mat{Z}}_i, \; \tilde{\Mat{Z}}_j) \Big\}^2 \label{eq:mds_constrained}
\end{align}
subject to the constraints that
\begin{align*}
    (i)~~\tilde{\Mat{Z}} = \begin{bmatrix} \Mat{Z} \Mat{W}^{(1)} \\ \vdots \\ \Mat{Z} \Mat{W}^{(M)} \end{bmatrix}, \; (ii)~ \Mat{W}^{(1)}, \ldots, \Mat{W}^{(M)} &\text{ are } p^* \times p^* \text{ diagonal matrices, and } 
    (iii)~\Mat{Z} \in \R^{n \times p^*}.
\end{align*}
Then the solution to this constrained MDS problem $\hat{\Mat{Z}}^*_{MDS}$ satisfies
\begin{align}
    \hat{\bZ}^{*}_{MDS} = 
    \begin{bmatrix} \hat{\bZ}^{*} \hat{\bW}^{(1)} \\ \vdots \\ \hat{\bZ}^{*} \hat{\bW}^{(M)} \end{bmatrix},
\end{align}
where $\hat \bZ^*$ and $\hat \bW^{(1)}, \ldots, \hat \bW^{(M)}$ are the solution to the CoMDS problem from (2) (from the main text). 

Hence, viewing (2) (from the main text) as a constrained MDS problem, we can solve (2) (from the main text) via an iterative MM algorithm called SMACOF \citep{deLeeuw2009}, which iterates between two main steps. The first step computes an update to the unconstrained MDS problem while the second step incorporates the constraint by projecting the unconstrained solution onto the feasible set. Specifically, after initializing $\tilde{\Mat{Z}} = \tilde{\Mat{Z}}^{(0)} \in \R^{Mn \times p^*}$, we repeat the two steps, detailed below, until convergence.

\paragraph{Step 1: Compute the unconstrained update $\bar{\Mat{Z}}^*$.}
Let $\mathcal{L}(\tilde{\Mat{Z}})$ denote the (unconstrained) stress function in \eqref{eq:mds_constrained} evaluated at some configuration $\tilde{\Mat{Z}} \in \R^{Mn \times p^*}$. Note that $\mathcal{L}$ can be rewritten as
\begin{align}
    \mathcal{L}(\tilde{\Mat{Z}})
    &:= \sum_{i < j} \Big\{ \Mat{D}_{ij} - d(\tilde{\Mat{Z}}_i, \; \tilde{\Mat{Z}}_j) \Big\}^2 \\
    &= \sum_{i < j} \bD_{ij}^{2} + \sum_{i < j} d^2({\tilde{\Mat{Z}}}_i, {\tilde{\Mat{Z}}}_j) - 2\sum_{i < j} \bD_{ij} d(\tilde{\Mat{Z}}_i, \tilde{\Mat{Z}}_j) \label{eq:loss_expanded}\\
    &= \text{constant} + \text{tr}\left(\tilde{\Mat{Z}}^{\top} \Mat{V} \tilde{\Mat{Z}} \right) - 2 \cdot \text{tr}\left(\tilde{\Mat{Z}}^{\top} \Mat{B}(\tilde{\Mat{Z}}) \tilde{\Mat{Z}}\right), \label{eq:loss_trace}
\end{align}
where the matrices $\Mat{V} \in \R^{Mn \times Mn}$ and $\Mat{B}(\tilde{\Mat{Z}}) \in \R^{Mn \times Mn}$ are defined with the entries:
\begin{align*}
    \Mat{V}_{ij} &= -1 \text{ if } i \neq j,
    &\Mat{B}(\tilde{\Mat{Z}})_{ij} &=
    \begin{cases}
        -\frac{\Mat{D}_{ij}}{d(\tilde{\Mat{Z}}_i, \tilde{\Mat{Z}}_j)} & \text{if } d(\tilde{\Mat{Z}}_i, \tilde{\Mat{Z}}_j) \neq 0 \text{ and } i \neq j\\
        0 & \text{if } d(\tilde{\Mat{Z}}_i, \tilde{\Mat{Z}}_j) = 0 \text{ and } i \neq j
    \end{cases},\\
    \Mat{V}_{ii} &= Mn-1,
    &\Mat{B}(\tilde{\Mat{Z}})_{ii} &= \sum_{j=1, j \neq i}^{Mn} \Mat{B}(\tilde{\Mat{Z}})_{ij}.
\end{align*}

Since $\mathcal{L}$ is non-convex and difficult to minimize directly, we instead construct a quadratic majorizing function $g$, which always upper bounds $\mathcal{L}$ and is easier to minimize. Specifically, by the Cauchy-Schwarz inequality, we have that for any $\Mat{U} \in \R^{Mn \times p^*}$,
\begin{align*}
    \mathcal{L}(\tilde{\Mat{Z}})
    &= \text{constant} + \text{tr}\left(\tilde{\Mat{Z}}^{\top} \Mat{V} \tilde{\Mat{Z}} \right) - 2 \text{tr}\left(\tilde{\Mat{Z}}^{\top} \Mat{B}(\tilde{\Mat{Z}}) \tilde{\Mat{Z}}\right) \\
    &\leq \text{constant} + \text{tr}\left(\tilde{\Mat{Z}}^{\top} \Mat{V} \tilde{\Mat{Z}} \right) - 2 \text{tr}\left(\tilde{\Mat{Z}}^{\top} \Mat{B}(\Mat{U}) \tilde{\Mat{U}}\right) =: g(\tilde{\Mat{Z}} \mid \Mat{U}).
\end{align*}

Now, the majorizing function $g(\tilde{\Mat{Z}} \mid \Mat{U})$ is a quadratic function of $\tilde{\Mat{Z}}$ and can be minimized directly by computing its derivative and setting it equal to zero:
\begin{align*}
    \frac{\partial g(\tilde{\Mat{Z}} \mid \Mat{U})}{\partial \tilde{\Mat{Z}}} = 2 \Mat{V} \tilde{\Mat{Z}} - 2 \Mat{B}(\Mat{U}) \Mat{U} = 0 \quad
    \implies \quad \Tilde{\Mat{Z}} = \Mat{V}^{+} \Mat{B}(\bU) \bU,
\end{align*}
where $\Mat{V}^{+}$ denotes the Moore–Penrose inverse of $\Mat{V}$. Thus, given the previous configuration $\tilde{\Mat{Z}}^{(t-1)}$, this results in the following unconstrained update, denoted as $\bar{\Mat{Z}}^{(t)}$ (also known as the Guttman transform \citep{guttman1968}):
\begin{align}
    \bar{\Mat{Z}}^{(t)} = \Mat{V}^{+} \Mat{B}(\tilde{\Mat{Z}}^{(t-1)}) \tilde{\Mat{Z}}^{(t-1)}. \label{eq:guttman}
\end{align}

\paragraph{Step 2: Incorporate the constraints.} To then incorporate the necessary constraint that $$\tilde{\Mat{Z}} = \begin{bmatrix} \Mat{Z} \Mat{W}^{(1)} \\ \vdots \\ \Mat{Z} \Mat{W}^{(M)} \end{bmatrix},$$ note that if $\tilde{\Mat{Z}}^{(t-1)}$ satisfies the imposed constraint, then as shown in \citet{deLeeuw1980} Theorem 1, the majorizing function can be rewritten as
\begin{align*}
    g(\tilde{\Mat{Z}} \mid {\tilde{\Mat{Z}}}^{(t-1)}) &= \text{constant} + \text{tr}\left((\tilde{\Mat{Z}} - \bar{\Mat{Z}}^{(t)})^\top \Mat{V} (\tilde{\Mat{Z}} - \bar{\Mat{Z}}^{(t)})\right) - \text{tr} \left(\bar{\Mat{Z}}^{(t)\top} \Mat{V} \bar{\Mat{Z}}^{(t)}\right).
\end{align*}

Thus, the constrained update $\tilde{\Mat{Z}}^{(t)}$ can be obtained by minimizing the right-hand side of the above equation over $\tilde{\Mat{Z}}$ subject to the given constraint. That is,
\begin{align}
    \tilde{\Mat{Z}}^{(t)} = \argmin_{\tilde{\Mat{Z}} \in \R^{Mn \times p^*}} \text{tr}\left((\tilde{\Mat{Z}} - \bar{\Mat{Z}}^{(t)})^\top \Mat{V} (\tilde{\Mat{Z}} - \bar{\Mat{Z}}^{(t)})\right) 
    \quad \text{subject to } \tilde{\Mat{Z}} = \begin{bmatrix} \Mat{Z} \Mat{W}^{(1)} \\ \vdots \\ \Mat{Z} \Mat{W}^{(M)} \end{bmatrix}, \label{eq:constrained_update}
\end{align}
which can be solved via alternating least squares (details in \citet{Commandeur} and \citet{Borg}). The two update steps \eqref{eq:guttman} and \eqref{eq:constrained_update} are then repeated until convergence.

\subsection{\localmethod{} Optimization Algorithm}\label{app:locomds_algorithm}

\localmethod{} can be solved using a similar MM algorithm as in \method{}, but with different constructions for $\Mat{V}$ and $\Mat{B}(\cdot)$. To see this, let $\delta_{ij}^{(m)}$ be an indicator of whether the sample pair $(i, j)$ is in the local neighborhood set $\mathcal{N}^{(m)}_{\pi}$ for dimension reduction method $m$, i.e.,
\begin{align*}
    \delta_{ij}^{(m)} = \begin{cases}
        1, & (i, j) \in \mathcal{N}^{(m)}_{\pi}\\
        0, & (i, j) \notin \mathcal{N}^{(m)}_{\pi}
    \end{cases}.
\end{align*}
Then, notice that the \localmethod{} stress function in (4) (from the main text) can be equivalently rewritten as
\begin{align*}
& \sum_{m = 1}^{M} \bigg\{ 
    \sum_{(i, j) \in \mathcal{N}^{(m)}_{\pi}} \left( \Mat{D}_{ij}^{(m)} - d(\Mat{W}^{(m)}\Mat{Z}_i, \Mat{W}^{(m)}\Mat{Z}_j) \right)^2
    - \; t_m \sum_{(i, j) \notin \mathcal{N}^{(m)}_{\pi}} d(\Mat{W}^{(m)}\Mat{Z}_i, \Mat{W}^{(m)}\Mat{Z}_j)
\bigg\} \\
&=\sum_{m = 1}^{M} \bigg\{ 
    \sum_{(i, j)} \delta_{ij}^{(m)} \left( \Mat{D}_{ij}^{(m)} - d(\Mat{W}^{(m)}\Mat{Z}_i, \Mat{W}^{(m)}\Mat{Z}_j) \right)^2
    - \; \sum_{(i, j)} t_m (1 - \delta_{ij}^{(m)}) d(\Mat{W}^{(m)}\Mat{Z}_i, \Mat{W}^{(m)}\Mat{Z}_j)
\bigg\} \\
&= \sum_{m = 1}^{M} \bigg\{ 
    \sum_{(i, j)} \delta_{ij}^{(m)} \left( \Mat{D}_{ij}^{(m)} \right)^2 
    + \sum_{(i, j)} \delta_{ij}^{(m)} d^2(\Mat{W}^{(m)}\Mat{Z}_i, \Mat{W}^{(m)}\Mat{Z}_j) \\
    &~~~~~~~~~~~~~~~~~ - 2 \sum_{(i, j)} \left(\delta_{ij}^{(m)} \Mat{D}_{ij}^{(m)} + t_m / 2 (1 - \delta_{ij}^{(m)})\right) d(\Mat{W}^{(m)}\Mat{Z}_i, \Mat{W}^{(m)}\Mat{Z}_j)
\bigg\},
\end{align*}
This takes a similar form as \eqref{eq:loss_expanded}.

Hence, similar to above, let us define the matrices $\Mat{V} \in \R^{Mn \times Mn}$ and $\Mat{B}(\tilde{\Mat{Z}}) \in \R^{Mn \times Mn}$ to be $\Mat{V} = \text{blockdiag}(\Mat{V}^{(1)}, \ldots, \Mat{V}^{(M)})$ and $\Mat{B}(\tilde{\Mat{Z}}) = \text{blockdiag}(\Mat{B}^{(1)}(\tilde{\Mat{Z}}), \ldots, \Mat{B}^{(M)}(\tilde{\Mat{Z}}))$, where the individual blocks $\Mat{V}^{(m)}$ and $\Mat{B}^{(m)}(\tilde{\Mat{Z}})$ have entries
\begin{align*}
    \Mat{V}^{(m)}_{ij} &= -\delta_{ij}^{(m)} \text{ if } i \neq j,
    &\Mat{B}^{(m)}(\tilde{\Mat{Z}})_{ij} &=
    \begin{cases}
        -\frac{\delta_{ij}^{(m)} \Mat{D}_{ij} + t_m/2(1 - \delta_{ij}^{(m)})}{d(\tilde{\Mat{Z}}_i, \tilde{\Mat{Z}}_j)} & \text{if } d(\tilde{\Mat{Z}}_i, \tilde{\Mat{Z}}_j) \neq 0 \text{ and } i \neq j\\
        0 & \text{if } d(\tilde{\Mat{Z}}_i, \tilde{\Mat{Z}}_j) = 0 \text{ and } i \neq j
    \end{cases},\\
    \Mat{V}^{(m)}_{ii} &= \sum_{j = 1, j \neq i}^{Mn} \delta_{ij}^{(m)},
    &\Mat{B}^{(m)}(\tilde{\Mat{Z}})_{ii} &= \sum_{j=1, j \neq i}^{Mn} \Mat{B}^{(m)}(\tilde{\Mat{Z}})_{ij}.
\end{align*}

Using this constructed $\Mat{V}$ and $\Mat{B}(\tilde{\Mat{Z}})$, it can be shown that the \localmethod{} stress function evaluated at some configuration $\tilde{\Mat{Z}} \in \R^{Mn \times p^*}$, denoted $\mathcal{L}_{\text{local}}(\tilde{\Mat{Z}})$, is equivalent to
\begin{align*}
    \mathcal{L}_{\text{local}}(\tilde{\Mat{Z}})
    &= \text{constant} + \text{tr}\left(\tilde{\Mat{Z}}^{\top} \Mat{V} \tilde{\Mat{Z}} \right) - 2 \cdot \text{tr}\left(\tilde{\Mat{Z}}^{\top} \Mat{B}(\tilde{\Mat{Z}}) \tilde{\Mat{Z}}\right),
\end{align*}
which is identical in form to \eqref{eq:loss_trace}. Consequently, \localmethod{} can be solved using the same iterative updates as \method{}, but with the above definitions for $\Mat{V}$ and $\Mat{B}(\cdot)$.

\section{Details on Empirical Studies}\label{app:empirical}
\subsection{Method Implementation}\label{app:methods}

Throughout this work, we use the same 16 dimension reduction methods as \citet{Ma-Meta} for input into each consensus dimension reduction method. The hyperparameters from each input dimension reduction method and each consensus dimension reduction method are summarized in Tables~\ref{tab:candidates_implementation} and \ref{tab:consensus_implementation}, respectively.

\begin{table}[h!]
\centering
\renewcommand{\arraystretch}{1.1}
\small
\rowcolors{2}{lightgray}{white}
\begin{tabular}{l p{7.5cm} c}
\toprule
\textbf{Method} & \textbf{Hyperparameters} & \textbf{\texttt{R} Package and Function} \\
\midrule
PCA & \texttt{k=2} & \texttt{rARPACK::svds}  \\
MDS & \texttt{k=2} & \texttt{stats::cmdscale}   \\
Non-metric MDS & \texttt{k=2}  & \texttt{MASS::isoMDS}   \\
Sammon & \texttt{k=2}  & \texttt{MASS::sammon}  \\
LLE & \texttt{m=2, k=20, reg=2}  & \texttt{lle::lle}   \\
HLLE & \texttt{method="HLLE", ndim=2, knn=20}  & \texttt{dimRed::embed}   \\
Isomap & \texttt{method="Isomap", ndim=2, knn=20}  & \texttt{dimRed::embed}  \\
kPCA & \texttt{method="kPCA", ndim=2, sigma=0.01} or \texttt{0.001}  & \texttt{dimRed::embed}  \\
LEIM & \texttt{method="LaplacianEigenmaps", ndim=2}  & \texttt{dimRed::embed}  \\
UMAP & \texttt{n\_components=2, n\_neighbors=30} or \texttt{50} & \texttt{uwot::umap}  \\
t-SNE & \texttt{method="tSNE", ndim=2, perplexity=30} or \texttt{50} & \texttt{dimRed::embed}  \\
PHATE  & \texttt{ndim=2, knn=30} or \texttt{50} & \texttt{phateR::phate}  \\
\bottomrule
\end{tabular}
\caption{Summary of input dimension reduction methods and the \texttt{R} functions used. Default hyperparameters are used unless otherwise specified.}
\label{tab:candidates_implementation}
\end{table}

\begin{table}[h!]
\centering
\renewcommand{\arraystretch}{1.1}
\small
\rowcolors{2}{lightgray}{white}
\begin{tabular}{l p{8.1cm} c}
\toprule
\textbf{Method} & \textbf{\centering Hyperparameters} & \textbf{\texttt{R} Package and Function} \\
\midrule
\method & \texttt{ndim=2, eps=1e-6, itmax=300} & \texttt{coMDS::coMDS} \\
\localmethod & \texttt{ndim=2, eps=1e-6, itmax=300, taus=c(10,5,1,0.5,0.1,0.05,0.01,0.005,0.001), percentiles=seq(0.1, 0.9, 0.1)} & \texttt{coMDS::locoMDS} \\
\metak & Gaussian kernel, \newline kernel bandwidth = median of meta-distances & N/A \\ 
\metau & \texttt{n\_components=2, n\_neighbors=30}  & \texttt{uwot::umap}\\
\msne & \texttt{k=2, perplexity=30} & \texttt{multiSNE::multiSNE} \\
\bottomrule
\end{tabular}
\caption{Summary of consensus dimension reduction methods and the hyperparameters used. Default hyperparameters are used unless otherwise specified. For Meta-Spec, R code was obtained from \url{https://github.com/rongstat/meta-visualization}.}
\label{tab:consensus_implementation}
\end{table}

\subsection{Datasets}\label{app:datasets}

Below, we provide more detailed descriptions of each dataset used in this study. Both raw and preprocessed datasets are available at \url{https://zenodo.org/records/17956327}.

\paragraph{Trajectory} This dataset describes the trajectory of the differentiation of mouse embryonic stem cells into primitive endoderm cells \citep{Hayashi-trajectory}. A total of 421 mouse embryonic stem cells were collected and individually recorded every 12 hours from the time of the induction of differentiation to 72 hours after the induction of differentiation. At each time point, approximately 70–90 cells were dissociated, isolated, and sequenced using the random displacement amplification sequencing (RamDA-seq) technique to obtain total RNA reads. The raw count data were preprocessed and normalized, resulting in a matrix of standardized expression levels for the 500 most variable genes across the 421 cells \citep {Ma-Meta}.

\paragraph{8EQ}
This dataset contains the Unique Molecular Identifier (UMI) counts of various cell types. The data was generated using the 10x Genomics GemCode protocol. A subset containing eight presorted subpopulations was selected. 
This dataset was preprocessed via PCA and the first 50 principal components were selected \citep{Huang-evaluation}. From the annotated cell types, eight subpopulations were then selected in unequal proportions, with each subgroup containing approximately 160–240 cells.

\paragraph{4EQ}
This dataset is derived from the same raw UMI count dataset described above \citep{Duo}, but represents a further subset of the eight presorted cell types used in the 8EQ dataset. Specifically, four cell types, B-cells, CD14 monocytes, naive cytotoxic T-cells and regulatory T-cells are selected from the eight available subpopulations.
This dataset was preprocessed via PCA and the first 100 principal components were selected \citep{Huang-evaluation}. We then randomly selected samples from these four subpopulations in equal proportions (300 cells per subpopulation).

\paragraph{HIV}
This dataset is a single-cell RNA sequencing (scRNA-seq) dataset, derived from peripheral blood mononuclear cells (PBMCs) collected from four individuals with HIV infection \citep{Kazer}. Transcriptional profiles of 59,286 cells in seven cell types were collected at multiple time points, from pre-infection through 1 year following viral detection. This dataset was preprocessed via PCA and the first 70 principal components were selected \citep{Huang-evaluation}. From the annotated cell types, seven subpopulations were then selected in equal proportions, with each subgroup containing 200 cells.

\paragraph{Olive Oil}
This dataset contains eight chemical measurements on the acid components for different samples of olive oils produced in nine different regions in Italy \citep{Forina}. This data was centered and scaled such that each feature had a mean of 0 and a standard deviation of 1.

\paragraph{Wheat}
This dataset contains measurements of geometric properties of wheat kernels from three distinct varieties: Kama, Rosa, and Canadian, with 70 samples per variety \citep{Charytanowicz-wheat}. Kernel images were obtained using a soft X-ray imaging technique, from which seven real-valued attributes were extracted using the GRAINS software package. This data was centered and scaled such that each feature had a mean of 0 and a standard deviation of 1.

\paragraph{Cycle}
This dataset is a mixture of cyclic structure and discrete clusters from the cell-division cycle of mouse embryonic stem cells \citep{Buettner-cycle}. There are 288 cells, with their cell-cycle stages (G1, S, and G2/M; 96 cells each) experimentally determined via flow-cytometry sorting. The cells were sequenced to obtain gene expression profiles, and the raw count data were then preprocessed and normalized. This resulted in a matrix of standardized expression levels for 1,147 related genes across the 288 cells \citep{Ma-Meta}.

\paragraph{Star}
This dataset contains a total of 10000 astronomical objects (stars, galaxies, and quasars) from the Sloan Digital Sky Survey (SDSS17) \citep{data-star}. For each astronomical object, recorded features include photometric measurements, astronomical coordinates, and important object identifiers. 
For this analysis however, we focused on a subset of 6 features that are most relevant for distinguishing between the type of astronomical object, that is, whether the object is a star, galaxy, or quasar.
These selected features include the ultraviolet, green, red, near-infrared, and infrared filters in the photometric system, as well as the redshift value based on the increase in wavelength.

We then randomly sampled 600 galaxies, 200 stars, and 200 quasars and centered and scaled such that each feature had a mean of 0 and a standard deviation of 1.

\paragraph{Wholesale}
This dataset contains information on 440 clients of a wholesale distributor, including whether each client is retail or hotel/cafe/restaurant, and the annual spending in monetary units across 6 product categories \citep{UCIWholesale}. Due to the highly right-skewed distribution of the spending amounts, we log-transformed the spending amounts prior to centering and scaling the data such that each (log-transformed) feature had a mean of 0 and a standard deviation of 1.

\section{Hyperparameter Tuning Details}\label{app:tuning}
\begin{figure}
    \centering
    \includegraphics[width=1.0\linewidth]{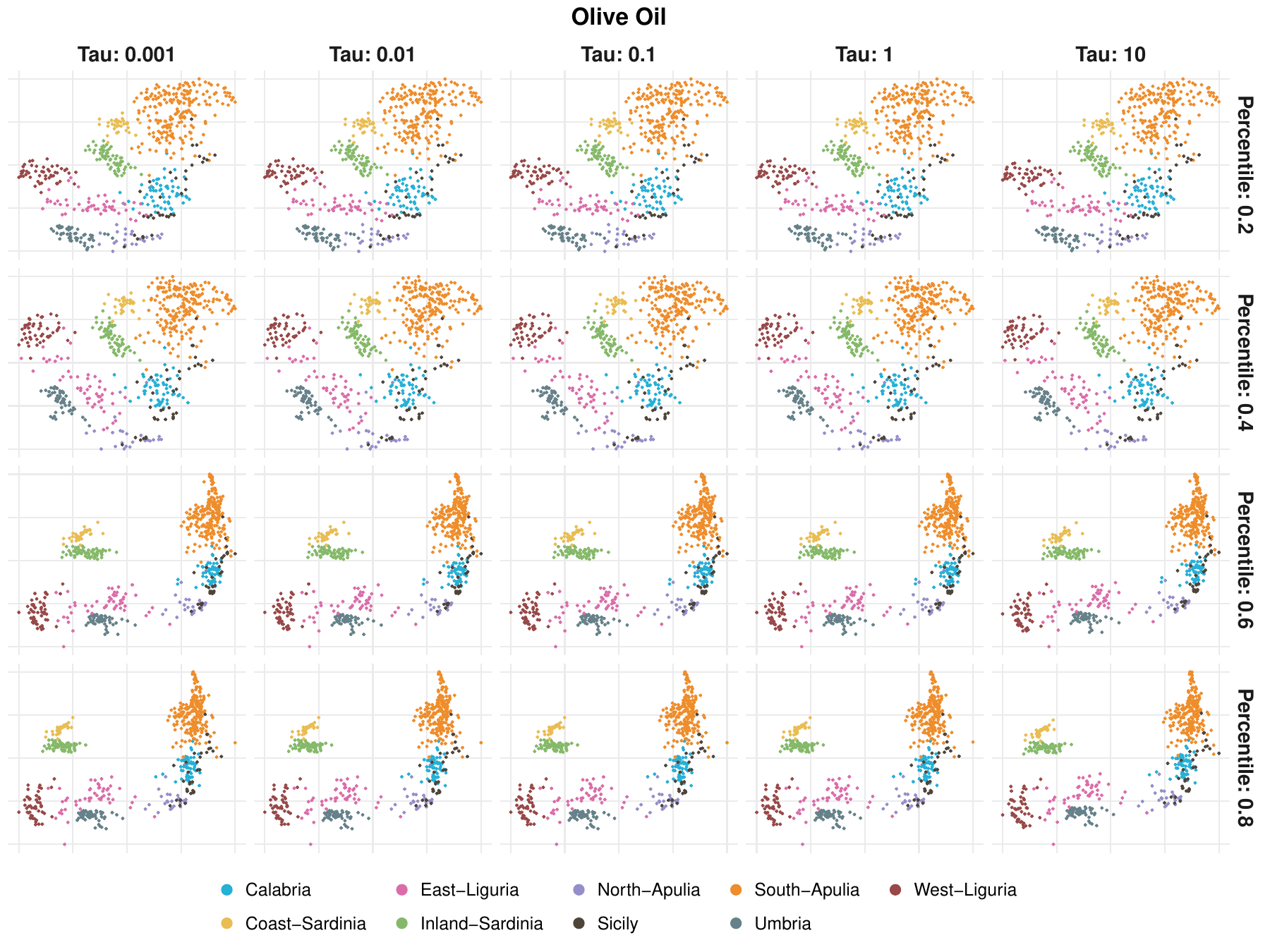}
    \caption{Low-dimensional embeddings obtained from \localmethod, under different combinations of hyperparameters ($\tau$, $\pi$), applied to the olive oil data. 
    }
    \label{fig:tuning_olive}
\end{figure}

\begin{figure}
    \centering
    \includegraphics[width=1.0\linewidth]{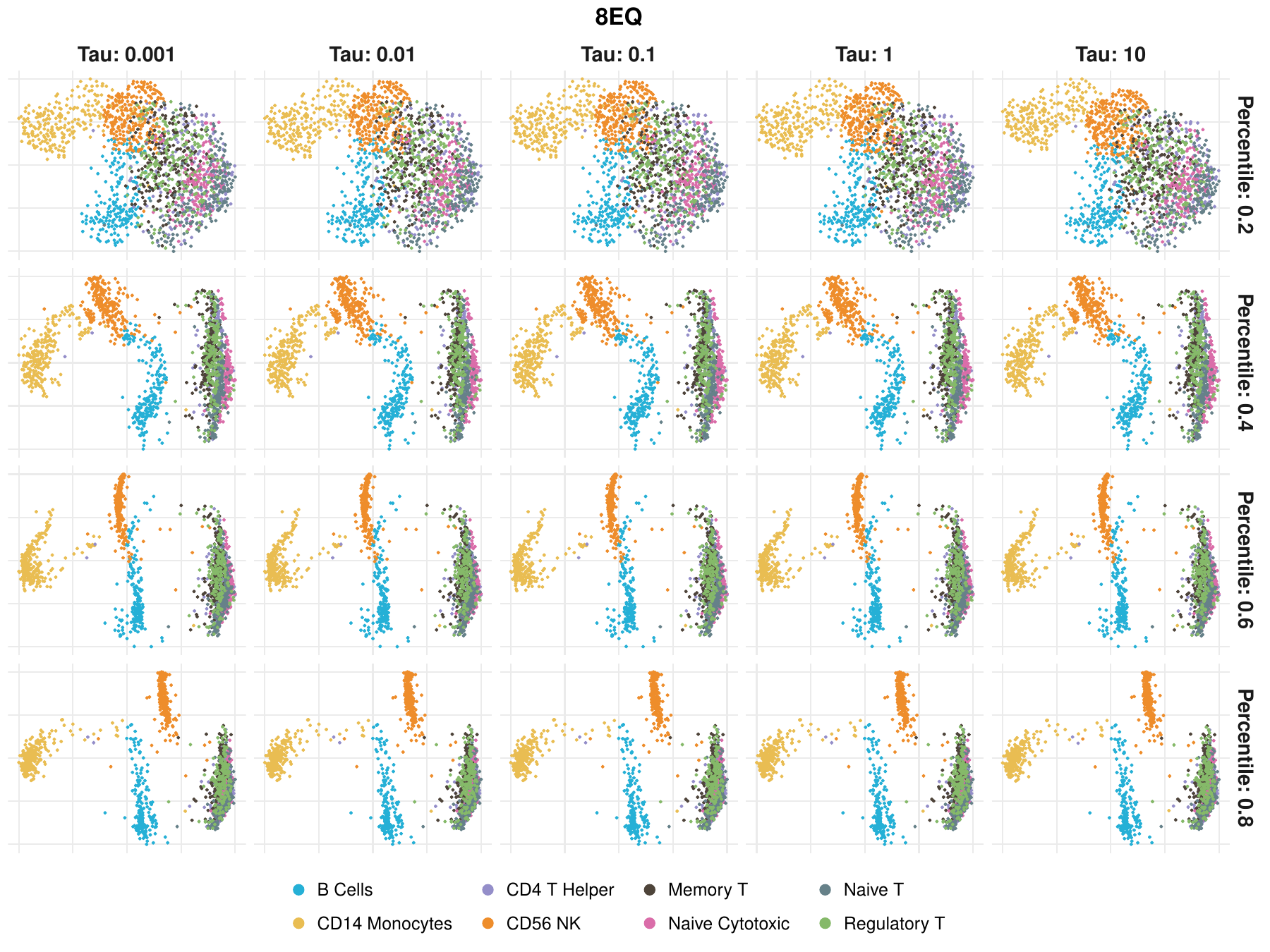}
    \caption{Low-dimensional embeddings obtained from \localmethod, under different combinations of hyperparameters ($\tau$, $\pi$), applied to the 8EQ data. 
    }
    \label{fig:tuning_8eq}
\end{figure}

To better understand the influence of the hyperparameters $\tau$ (repulsion strength) and $\pi$ (neighborhood size) on \localmethod{}, we present the \localmethod{} embeddings obtained from a large grid of $(\tau, \pi)$ values on two representative datasets, olive oil and 8EQ, in Figures~\ref{fig:tuning_olive} and \ref{fig:tuning_8eq}, respectively. 

As expected, $\pi$ controls the degree of locality in the embeddings, with larger values of $\pi$ resulting in \localmethod{} embeddings that grow increasingly similar to \method{} (refer to Appendix~\ref{app:plots} for comparison).
On the other hand, for a fixed percentile $\pi$, the \localmethod{} embeddings appear to be relatively stable across a wide range of $\tau$ values.
While Figures~\ref{fig:tuning_olive} and \ref{fig:tuning_8eq} only show the results for two datasets here, we observed similar patterns across all nine datasets considered in this study.
This may suggest that in practice, one can allocate more computational resources to tuning $\pi$ across a finer grid while using a coarser grid (or potentially even a single value) for $\tau$.
For all empirical studies in this work however, we went ahead and tuned both $\pi$ and $\tau$, each over a grid of 9 values detailed in Table~\ref{tab:consensus_implementation}.

\begin{figure}
    \centering
    \includegraphics[width=1.0\linewidth]{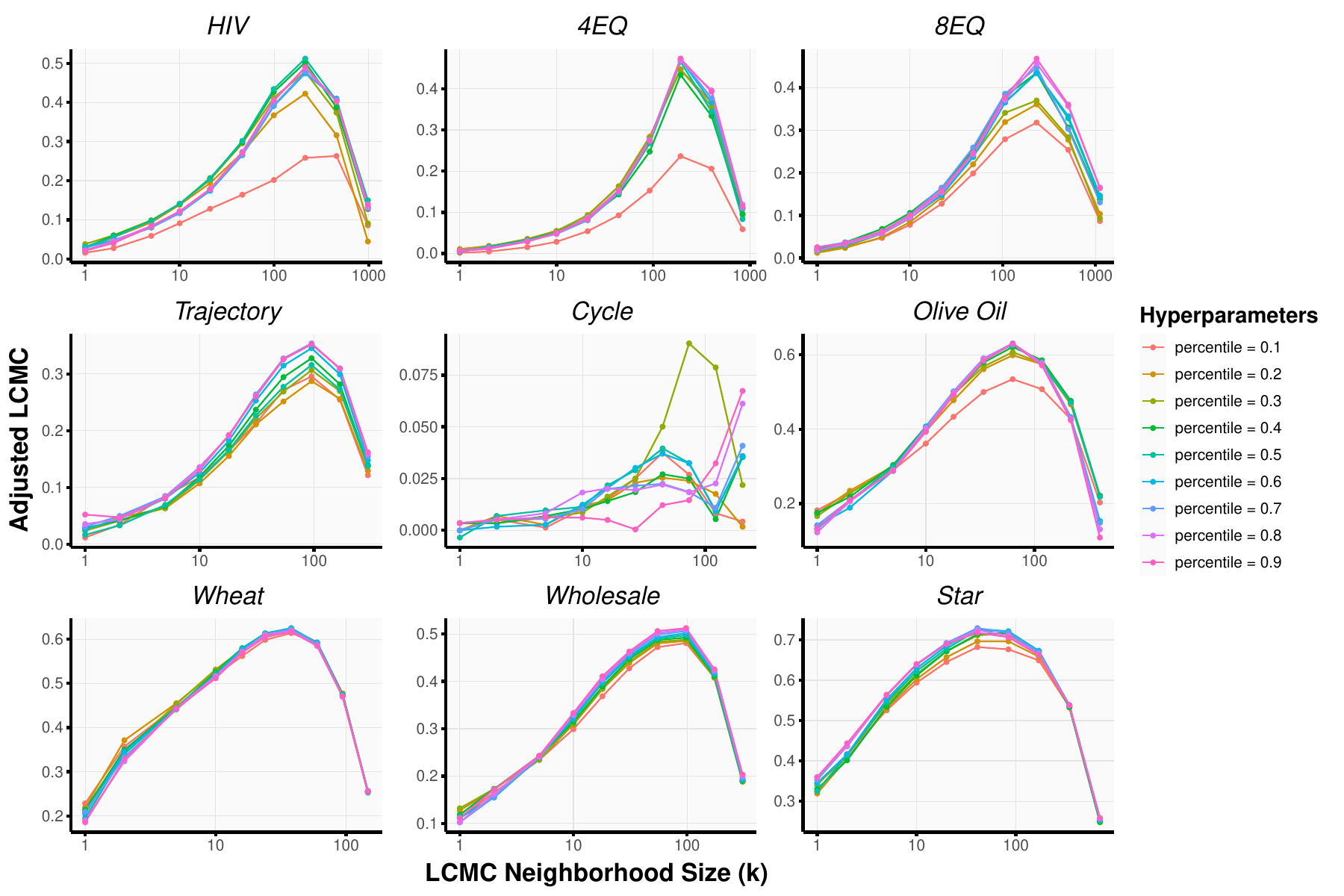}
    \caption{Adjusted LCMC scores across different percentiles $\pi$ and the size ($k$) of the neighborhood in the adjusted LCMC computation for each real-world dataset used in this study.}
    \label{fig:hyperparameter_tuning}
\end{figure}

To then select the best hyperparameters $(\tau, \pi)$, we computed the adjusted LCMC scores across all considered combinations of ($\tau$, $\pi$).
Figure~\ref{fig:hyperparameter_tuning} visualizes the adjusted LCMC scores across varying percentiles $\pi$ for all nine real-world datasets used in this work.
For clarity, we omit the adjusted LCMC curves for different $\tau$ values in Figure~\ref{fig:hyperparameter_tuning} since they were largely similar for a fixed $\pi$ (as also depicted in Figures~\ref{fig:tuning_olive} and \ref{fig:tuning_8eq}).
We further note that the peak of the adjusted LCMC curves tended to match the average cluster size in each dataset, motivating our recommendation to select the pair ($\tau$, $\pi$) that maximizes the adjusted LCMC score across $k$, up to the point at which the adjusted LCMC score begins to decrease (detailed in Section 2.2 in the main text).

\newpage
\section{Additional Empirical Results}\label{app:results}
In this section, we provide additional results to supplement those presented in Section 4 in the main text, regarding the evaluation of (i) global and local structure preservation, (ii) supervised prediction performance, and (iii) the stability of the proposed methods.

\subsection{Additional Global and Local Structure Preservation Results}\label{app:eval_structure}

First, to help summarize the global and local structure preservation results across all nine datasets (originally presented in Figure 6 in the main text), we show the number of datasets (out of 9) for which each method performed the best (i.e., rank 1), second best (i.e., rank 2), and so on, according to each global or local structure preservation metric in Table~\ref{tab:method_ranks_structure}. 
The average rank for each method across all datasets is also provided in the final column.
These results indicate that \method{}, followed closely by \localmethod{}, consistently outperform the other consensus dimension reduction approaches in terms of preserving the overall global structure of the data.
Moreover, \localmethod{} achieves the best average rank for local LCMC, indicating its strong performance in also preserving the local structure of the data.

\begin{table}[h!]
\centering
\renewcommand{\arraystretch}{1.2}
\small
\begin{tabular}{l c c c c c c}
\toprule
\multicolumn{7}{c}{{\textit{(A)}} \textbf{Random Triplet Accuracy}} \\
\textbf{Method} & Rank 1 & Rank 2 & Rank 3 & Rank 4 & Rank 5 & \textbf{Average Rank}\\
\midrule
\method & 6 & 3 & 0 & 0 & 0 & \textbf{1.33} \\
\localmethod & 3 & 5 & 0 & 1 & 0 & 1.89 \\
\metau & 0 & 1 & 3 & 3 & 2 & 3.67 \\
\metak & 0 & 0 & 3 & 2 & 4 & 4.11 \\
\msne & 0 & 0 & 3 & 3 & 3 & 4.00 \\
\bottomrule
\end{tabular}

\vspace{0.3cm}

\begin{tabular}{l c c c c c c}
\toprule
\multicolumn{7}{c}{{\textit{(B)}} \textbf{Spearman correlation}} \\
\textbf{Method} & Rank 1 & Rank 2 & Rank 3 & Rank 4 & Rank 5 & \textbf{Average Rank}\\
\midrule
\method & 6 & 2 & 1 & 0 & 0 & \textbf{1.44} \\
\localmethod & 3 & 5 & 1 & 0 & 0 & 1.77 \\
\metau & 0 & 0 & 2 & 5 & 2 & 4.00 \\
\metak & 0 & 2 & 2 & 2 & 3 & 3.67 \\
\msne & 0 & 0 & 3 & 2 & 4 & 4.11 \\
\bottomrule
\end{tabular}

\vspace{0.3cm}

\begin{tabular}{l c c c c c c}
\toprule
\multicolumn{7}{c}{{\textit{(C)}} \textbf{Local LCMC}} \\
\textbf{Method} & Rank 1 & Rank 2 & Rank 3 & Rank 4 & Rank 5 & \textbf{Average Rank}\\
\midrule
\method & 1 & 2 & 2 & 3 & 1 & 3.11 \\
\localmethod & 3 & 3 & 2 & 1 & 0 & \textbf{2.11}
\\
\metau & 0 & 3 & 3 & 3 & 0 & 3.00 \\
\metak & 1 & 0 & 0 & 0 & 8 & 4.56 \\
\msne & 4 & 1 & 2 & 2 & 0 & 2.22 \\
\bottomrule
\end{tabular}

\caption{
    Number of datasets (out of 9) for which each method performed the best (i.e., rank 1), second-best (i.e., rank 2), etc, according to (A) the random triplet accuracy, (B) the Spearman correlation, and (C) the local LCMC metric, as defined in Section 4.2 in the main text. Rank 1 corresponds to the best-performing method for a given dataset while Rank 5 corresponds to the worst-performing method.
}

\label{tab:method_ranks_structure}
\end{table}

\begin{figure}[H]
    \centering
    \includegraphics[width=1.0\linewidth]{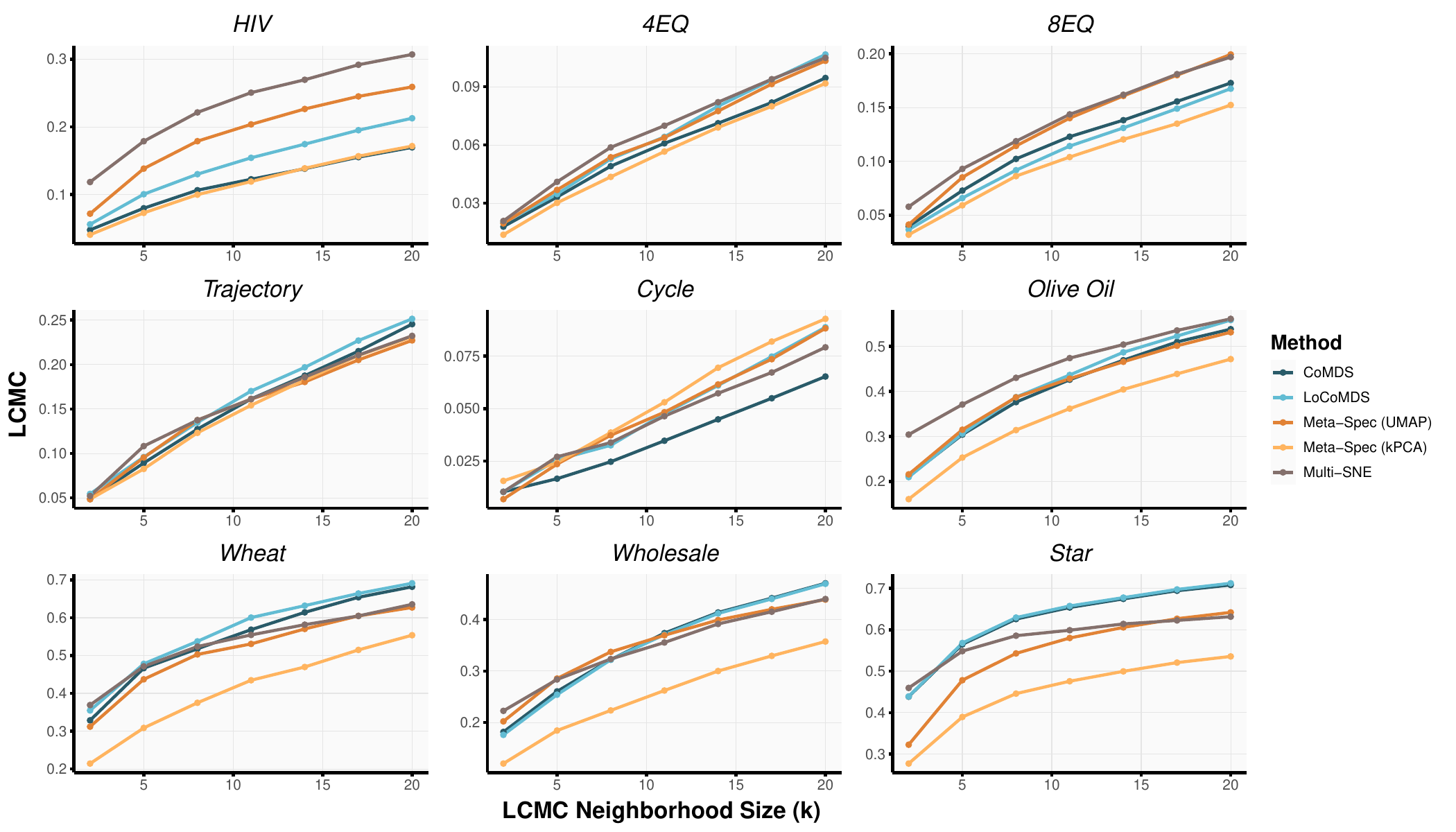}
    \caption{Local Continuity Meta-Criterion (LCMC) scores across nine datasets for different consensus dimensionality reduction methods as a function of the number of nearest neighbors $k$.}
    \label{fig:lcmc_curve}
\end{figure}

Recall though that the local LCMC results in Figure 6 in the main text and Table~\ref{tab:method_ranks_structure} were averaged across different values of $k$ (the number of nearest neighbors in the LCMC calculation) in order to simplify the comparison. 
We thus also provide the full LCMC curves in Figure~\ref{fig:lcmc_curve}. Here, for each consensus dimension reduction method, we show the local LCMC value across all considered values of $k$, which ranges from 2 to 20 with a step size of 3.
Generally, the choice of $k$ does not appear to drastically change the relative performance of the different methods, and the averaged results in Figure 6 in the main text and Table~\ref{tab:method_ranks_structure} are representative of the overall trends observed across different values of $k$.

\subsection{Additional Supervised Evaluation and Prediction Accuracy Results}

\begin{figure}
    \centering
    \includegraphics[width=1\linewidth]{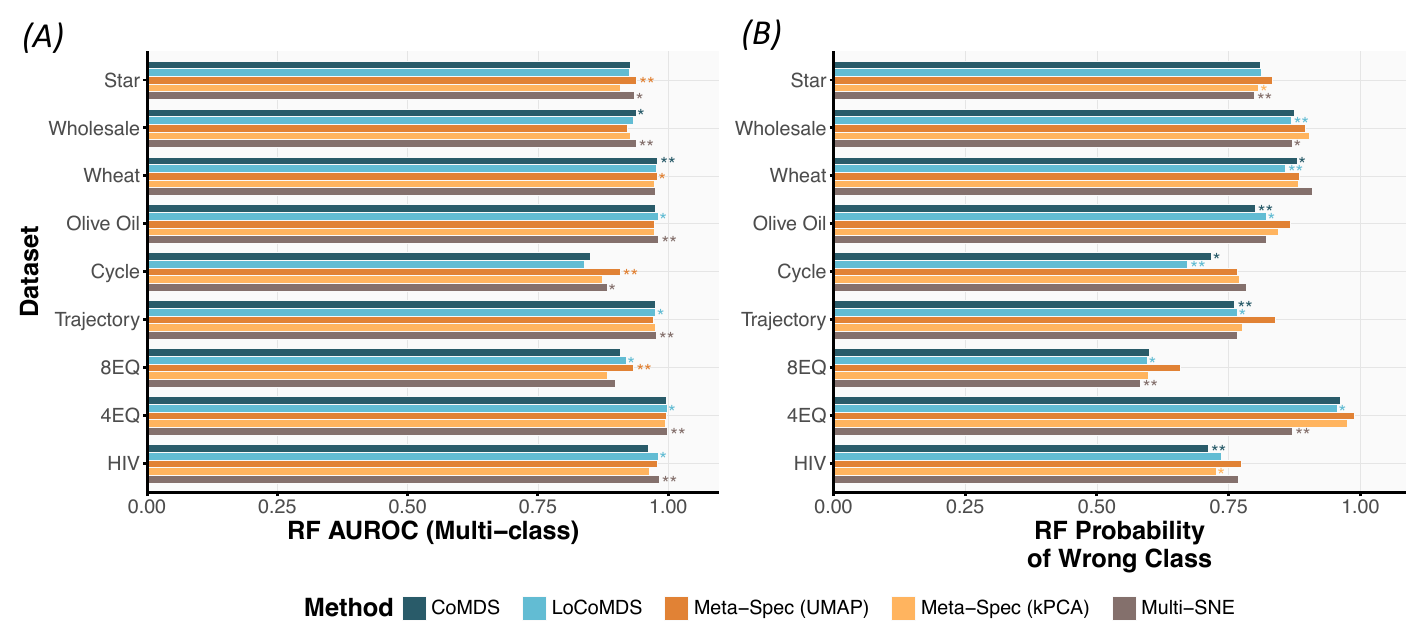}
    \caption{
        (A) Random forest AUROC results across all nine datasets. Higher values indicate better performance. (B) The average RF predicted probability of the wrong class for misclassified samples across all nine datasets. Lower values indicate better performance. For each dataset and metric, the method with the best performance is denoted by two asterisks while the second-best method is denoted by one asterisk.
    }
    \label{fig:rf_eval}
\end{figure}

In Figure~\ref{fig:rf_eval}, we provide the full supervised evaluation results complementing those presented in Table 2 in Section 4.2.2 in the main text. Specifically, for each dataset, we show the test AUROC from the random forest (RF) that was trained to predict the accompanying cluster labels. We also show the average predicted probability of the predicted (but wrong) class for misclassified samples. Recall high predicted probabilities for the wrong class indicate that the RF was overly-confident in its incorrect predictions. Existing methods tend to lead to more confident, but incorrect predictions whereas \method{} and \localmethod{} tend to exhibit more uncertainty in their incorrect predictions while still achieving high overall prediction accuracy.

\subsection{Additional Stability Case Study Results}\label{app:stability}

\paragraph{Stability to Method Choice.} 
In Section 4.2.3, we observed that the consensus embeddings from \localmethod{}, applied to the 8EQ dataset using two different sets of input dimension reduction methods, were highly similar.
There, the two sets of input methods consisted of (i) two base methods (i.e., the two input dimension reduction methods with the highest eigenscores \citep{Ma-Meta}) and (ii) those two base methods augmented with HLLE.

\begin{figure}[h]
    \centering
    \includegraphics[width=1.0\linewidth]{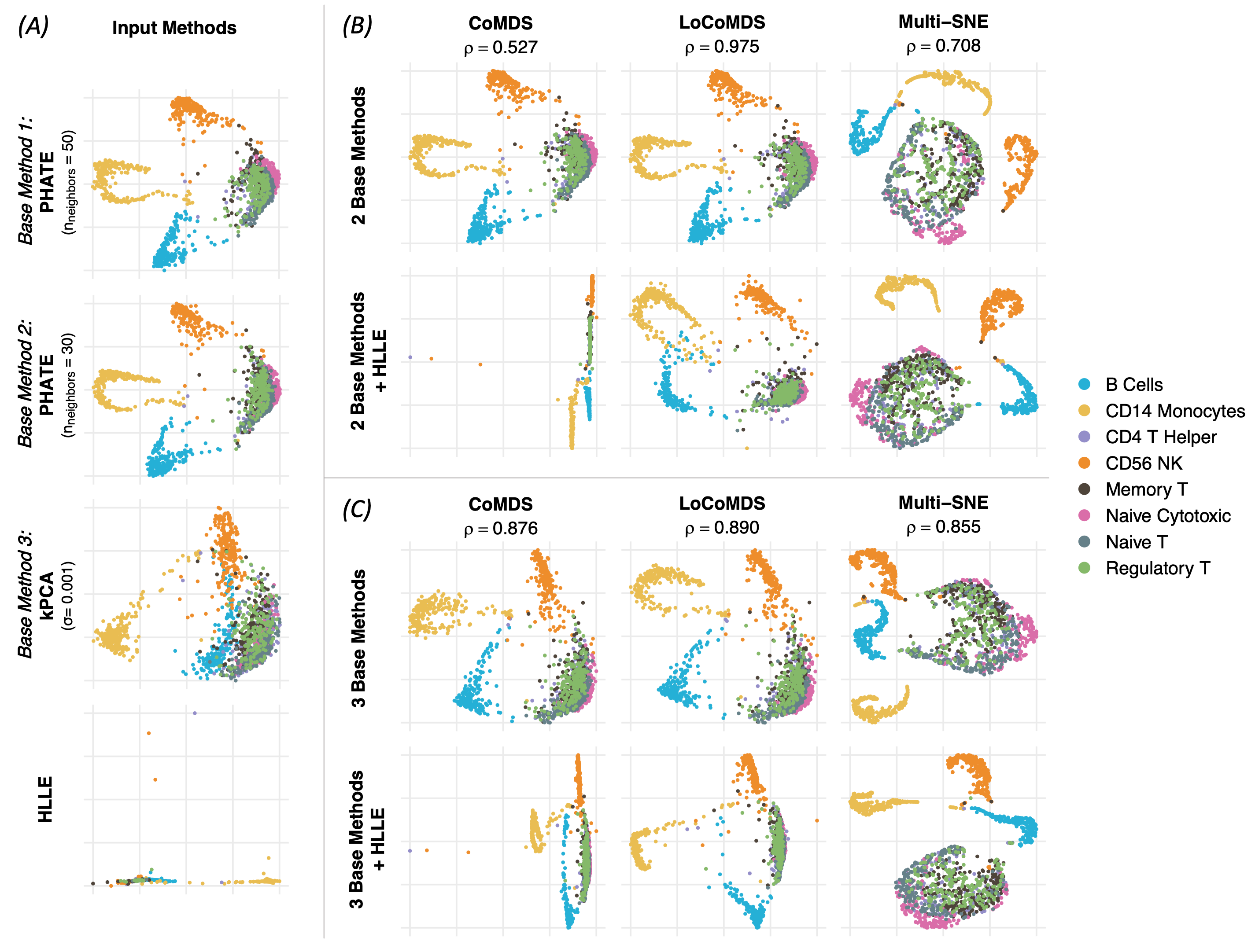}
    \caption{
        (A)~Low-dimensional embeddings from three base methods (i.e., PHATE with number of neighbors equal to 30 or 50 and kPCA with bandwidth $\sigma = 0.001$) as well as HLLE on the 8EQ dataset. 
        (B)~Low-dimensional embeddings obtained from \method{}, \localmethod{}, and \msne{} on the 8EQ dataset using (top) the first two base methods as input, compared to (bottom) the same two base methods plus HLLE as input. 
        (C)~Same as (B) but using all three base methods as input.
        For each method, the Mantel test statistic \citep{Mantel}, measuring the correlation between the two embeddings with and without HLLE, is reported and denoted by $\rho$.
    }
    \label{fig:method_stability_8eq_all}
\end{figure}

\begin{figure}[h]
    \centering
    \includegraphics[width=1.0\linewidth]{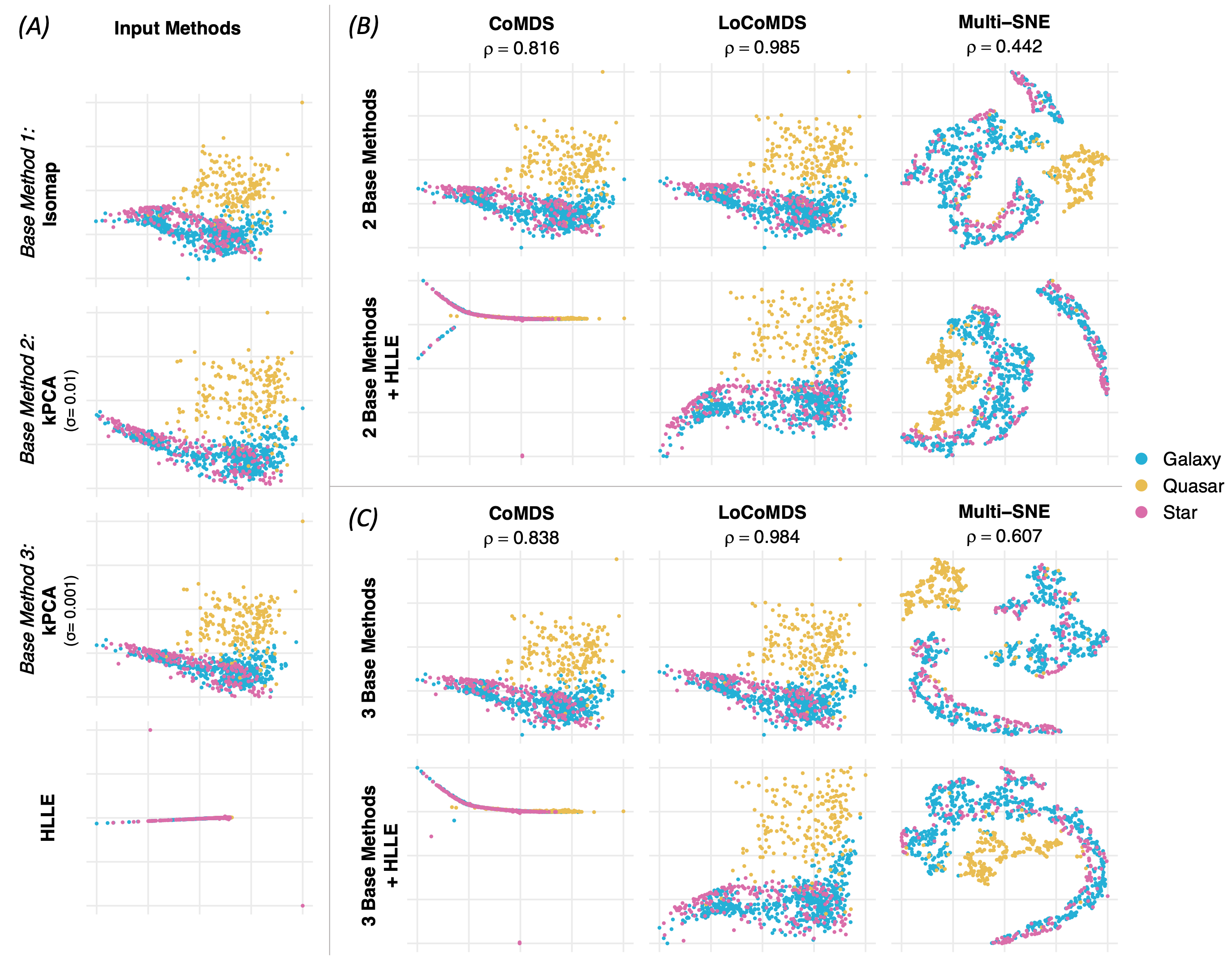}
    \caption{
        (A)~Low-dimensional embeddings from three base methods (i.e., Isomap and kPCA with bandwidth $\sigma = 0.01$ or $0.001$) as well as HLLE on the star dataset. 
        (B)~Low-dimensional embeddings obtained from \method{}, \localmethod{}, and \msne{} on the star dataset using (top) the first two base methods as input, compared to (bottom) the same two base methods plus HLLE as input. (C)~Same as (B) but using all three base methods as input.
        For each method, the Mantel test statistic \citep{Mantel}, measuring the correlation between the two embeddings with and without HLLE, are reported and denoted by $\rho$.
    }
    \label{fig:method_stability_star_all}
\end{figure}

To further demonstrate the stability (or instability) of \localmethod{} and other consensus dimension reduction methods to the choice of input methods, we expanded this analysis to include another dataset (i.e., the star dataset) and larger numbers of input/base methods (see Figures~\ref{fig:method_stability_8eq_all} and \ref{fig:method_stability_star_all}). 
Specifically, instead of using only two base methods, we also considered the case of using three base methods, analogously defined as the top three input dimension reduction methods according to their eigenscores \citep{Ma-Meta}.

These additional results in Figures~\ref{fig:method_stability_8eq_all} and \ref{fig:method_stability_star_all} corroborate the main takeaways from Section 4.2.3 in the main text. 
Namely, first, when using only the base methods as inputs, \method{} and \localmethod{} output visualizations that closely resemble these inputs since the base methods are themselves all highly similar to each other. In contrast, \msne{} introduces its own idiosyncracies and produces a noticeably different visualization from the inputs.
Second, when including both the base methods and HLLE as input, \method{} tends to be heavily impacted by the outliers, though this impact is lessened when including three base methods instead of two.
Lastly, \localmethod{} continues to largely reflect the common structure shared among the base methods, regardless of whether or not the HLLE is included as an input method.
This demonstrated stability generally helps to reduce the burden of method selection by the practitioner.

\paragraph{Stability to Hyperparameter Choice.}

To lastly illustrate another example of how our consensus dimension reduction framework can be used to address the instability of existing dimension reduction methods to hyperparameter choice, we repeated the analysis from Section 4.2.3 in the main text on the HIV data. 

\begin{figure}
    \centering
    \includegraphics[width=0.9\linewidth]{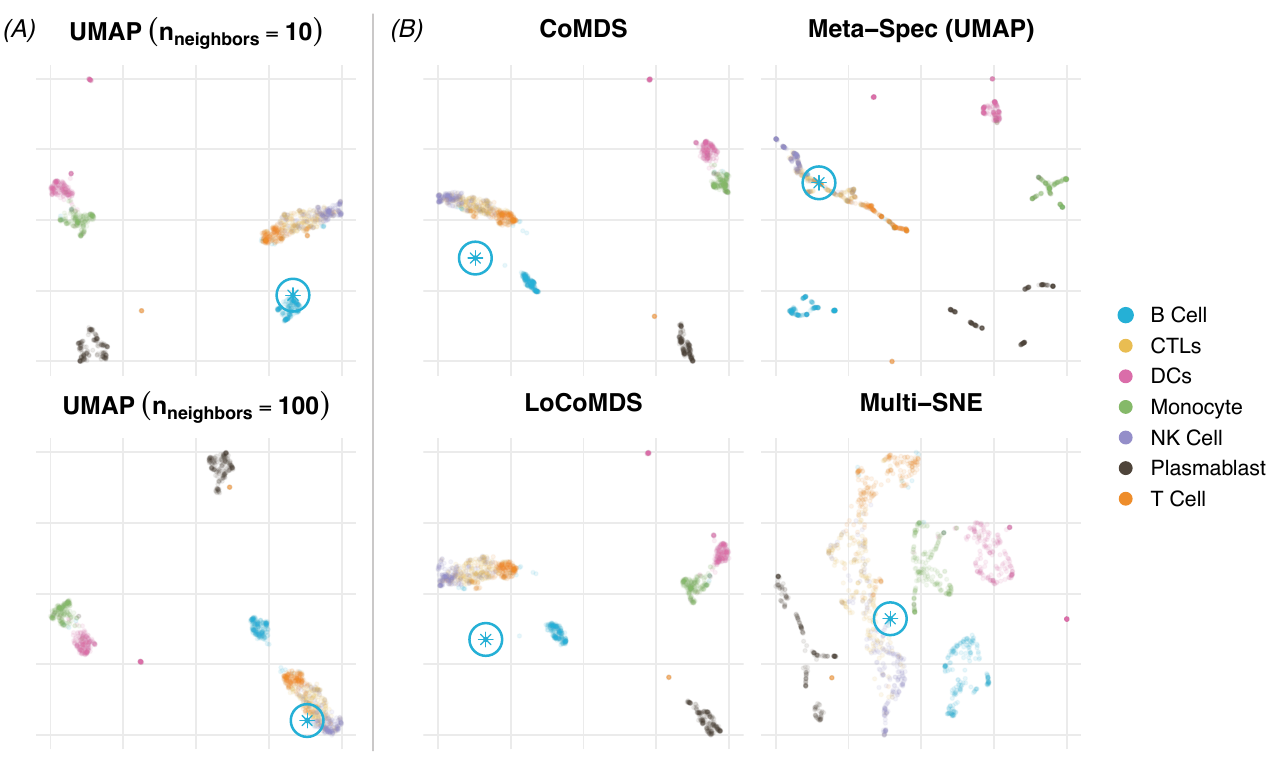}
    \caption{(A) Low dimensional visualizations obtained from UMAP with the number of neighbors of 10 and 100 on the HIV data. (B) Low-dimensional visualizations obtained from consensus dimension reduction methods using these two UMAP representations as input. The point circled and marked with an asterisk was assigned to noticeably different clusters under the two UMAP hyperparameter settings.}
    \label{fig:stability_hiv}
\end{figure}

As seen in Figure~\ref{fig:stability_hiv}A, UMAP gives different results for different choices of the number of neighbors hyperparameter.
In particular, there is a B cell (circled and marked with an asterisk) that is placed in the B cell cluster when using 10 nearest neighbors but near the CTLs and NK cells when using 100 nearest neighbors.
When using the two UMAP embeddings as input, both \method{} and \localmethod{} put this particular B cell between the two clusters (i.e., between the B cells and the CTLs/NK cells), reflecting the uncertainty in its position (Figure~\ref{fig:stability_hiv}B). 
This is unlike the existing methods, which fail to capture this uncertainty in their consensus embeddings.

\clearpage
\section{Full Dimension Reduction Plots}\label{app:plots}
For completeness, we lastly provide the full set of low-dimensional visualizations, obtained from all 16 input dimension reduction methods as well as from all consensus dimension reduction methods (\method{}, \localmethod{}, Meta-Spec (kPCA), Meta-Spec (UMAP), and Multi-SNE) for each simulated and real-world dataset used in this study.

\begin{figure}[H]
    \centering
    \includegraphics[width=1.0\linewidth]{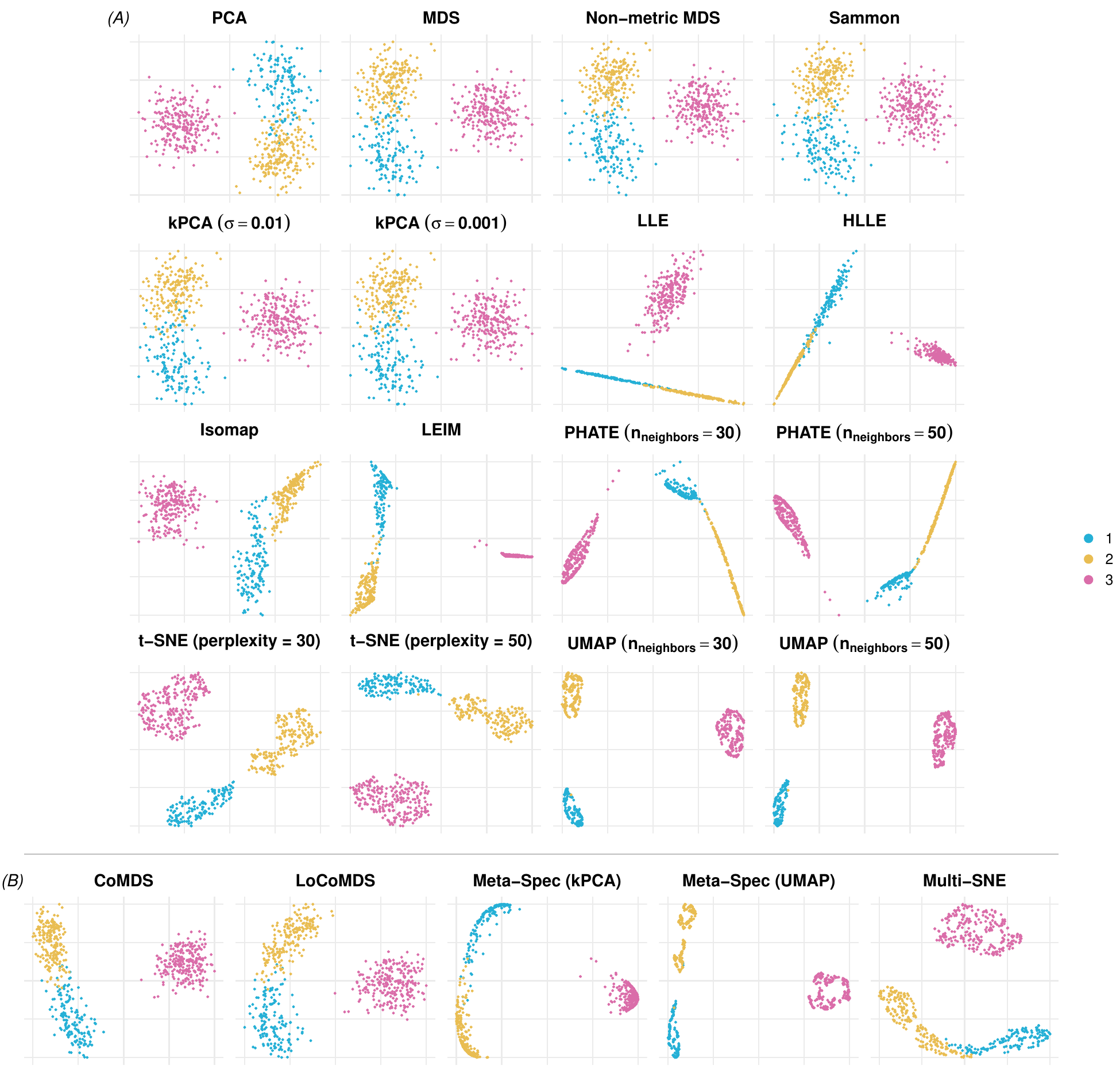}
    \caption{Low-dimensional embeddings, obtained from (A) all 16 input dimension reduction methods and (B) all consensus dimension reduction methods under consideration, applied to the \textbf{simulated mixture of Gaussians data} from Section 3.1 in the main text.}
    \label{fig:full_gaussian}
\end{figure}

\begin{figure}
    \centering
    \includegraphics[width=1.0\linewidth]{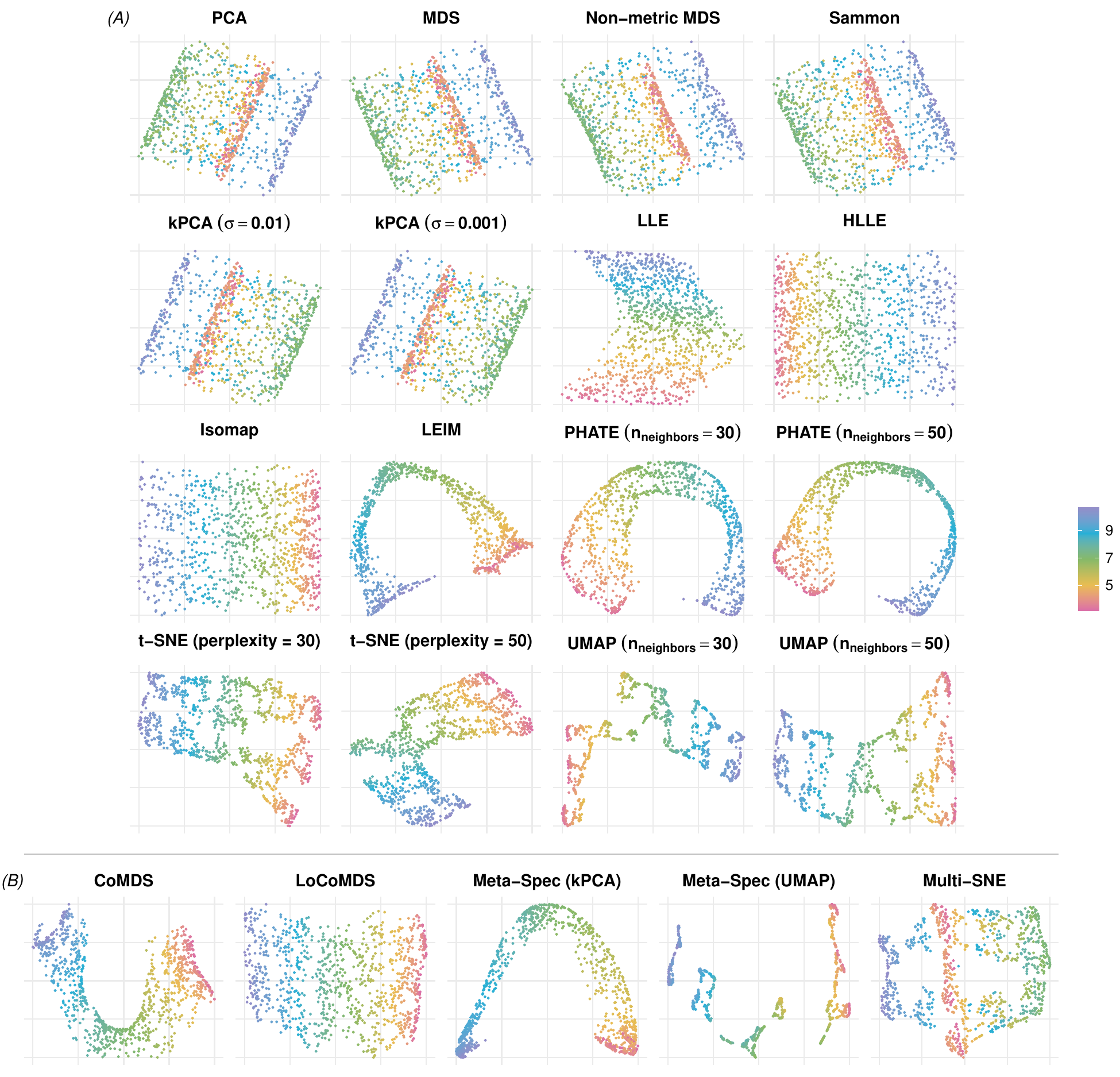}
    \caption{Low-dimensional embeddings, obtained from (A) all 16 input dimension reduction methods and (B) all consensus dimension reduction methods under consideration, applied to the \textbf{simulated Swiss roll data} from Section 3.2 in the main text.}
    \label{fig:full_swiss}
\end{figure}

\begin{figure}
    \centering
    \includegraphics[width=1.0\linewidth]{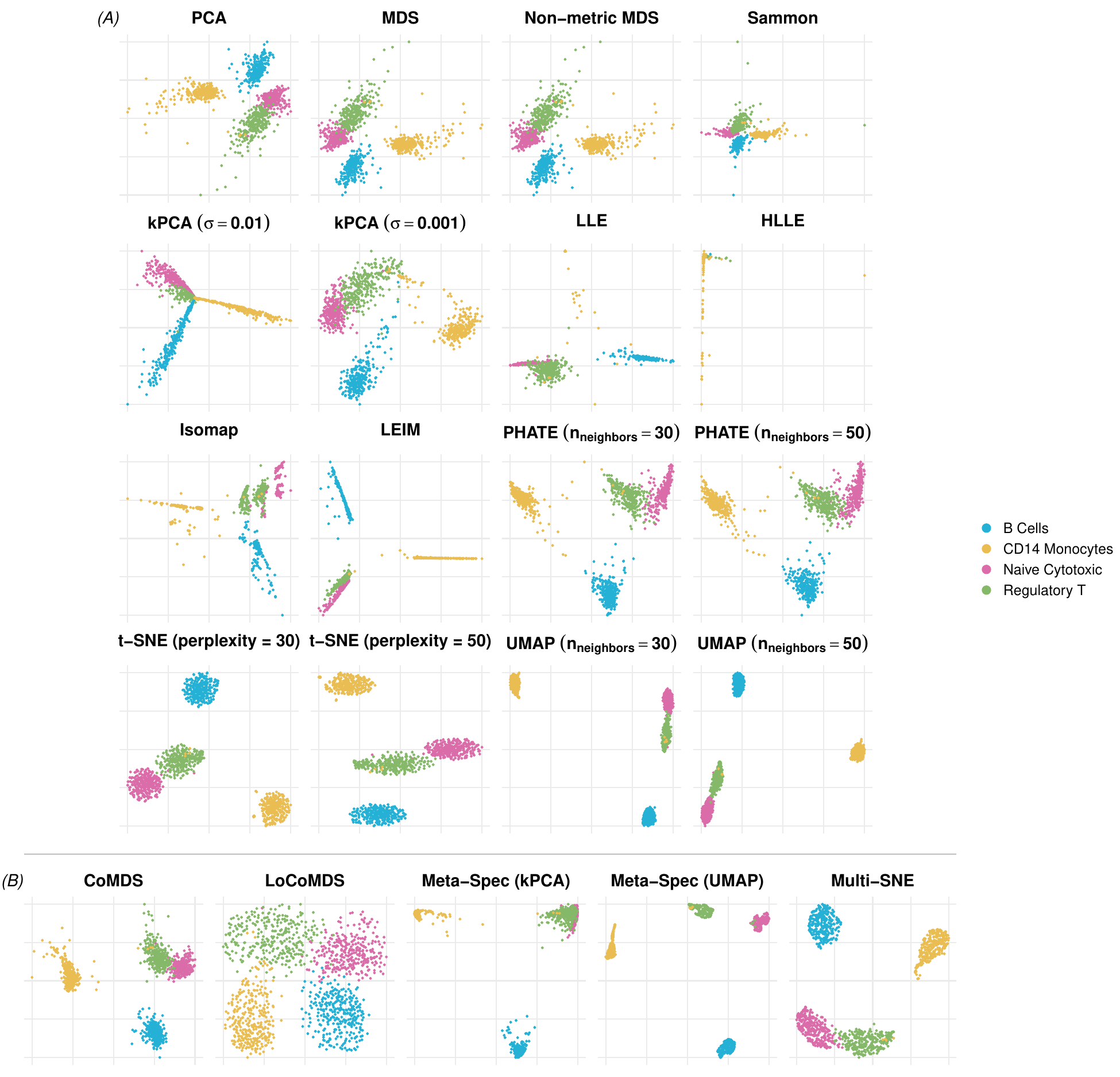}
    \caption{Low-dimensional embeddings, obtained from (A) all 16 input dimension reduction methods and (B) all consensus dimension reduction methods under consideration, applied to the \textbf{4EQ data}.}
    \label{fig:full_4eq}
\end{figure}

\begin{figure}
    \centering
    \includegraphics[width=1.0\linewidth]{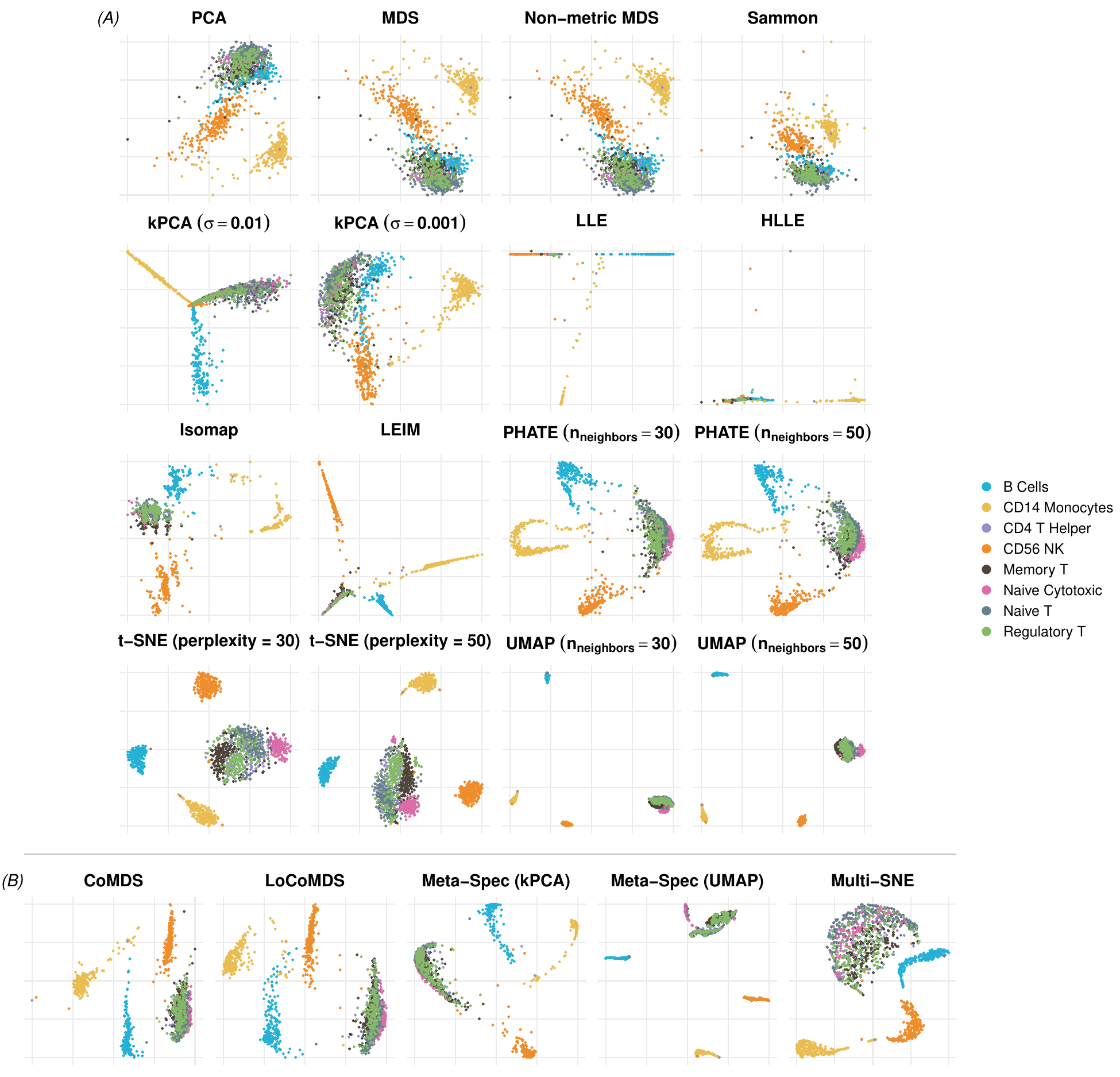}
    \caption{Low-dimensional embeddings, obtained from (A) all 16 input dimension reduction methods and (B) all consensus dimension reduction methods under consideration, applied to the \textbf{8EQ data}.}
    \label{fig:full_8eq}
\end{figure}

\begin{figure}
    \centering
    \includegraphics[width=1.0\linewidth]{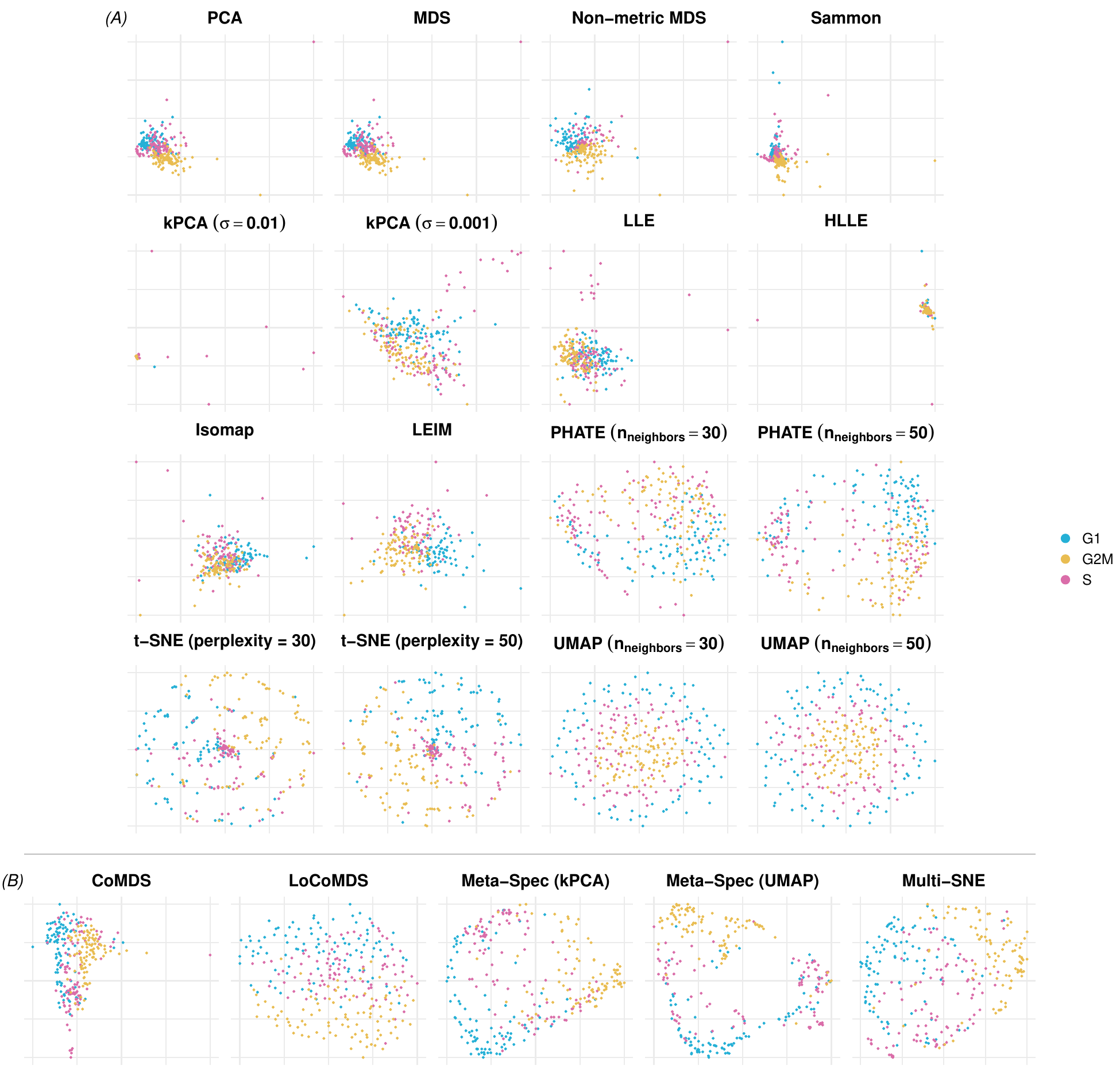}
    \caption{Low-dimensional embeddings, obtained from (A) all 16 input dimension reduction methods and (B) all consensus dimension reduction methods under consideration, applied to the \textbf{cycle data}.}
    \label{fig:full_cycle}
\end{figure}

\begin{figure}
    \centering
    \includegraphics[width=1.0\linewidth]{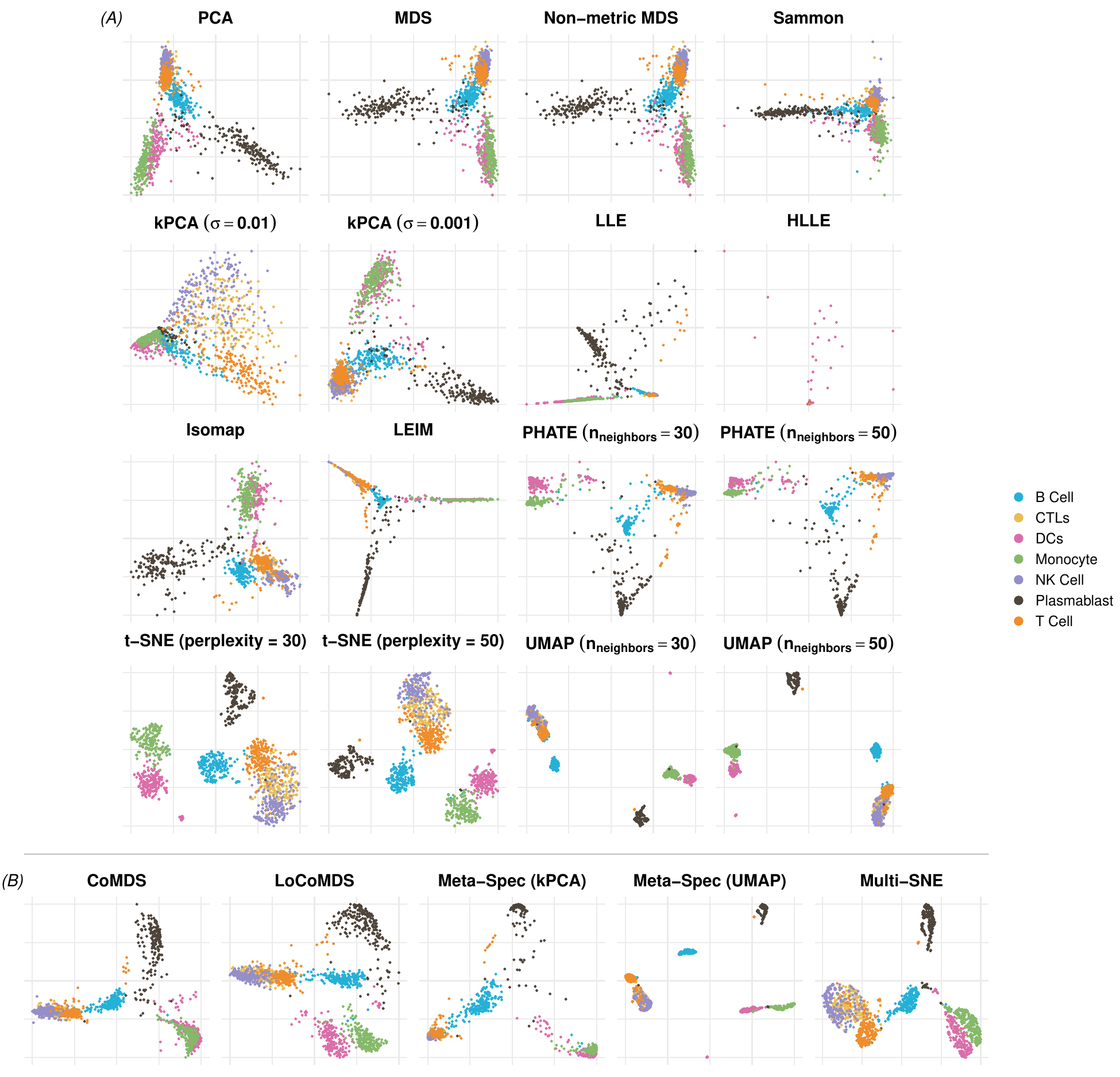}
    \caption{Low-dimensional embeddings, obtained from (A) all 16 input dimension reduction methods and (B) all consensus dimension reduction methods under consideration, applied to the \textbf{HIV data}.}
    \label{fig:full_hiv}
\end{figure}

\begin{figure}
    \centering
    \includegraphics[width=1.0\linewidth]{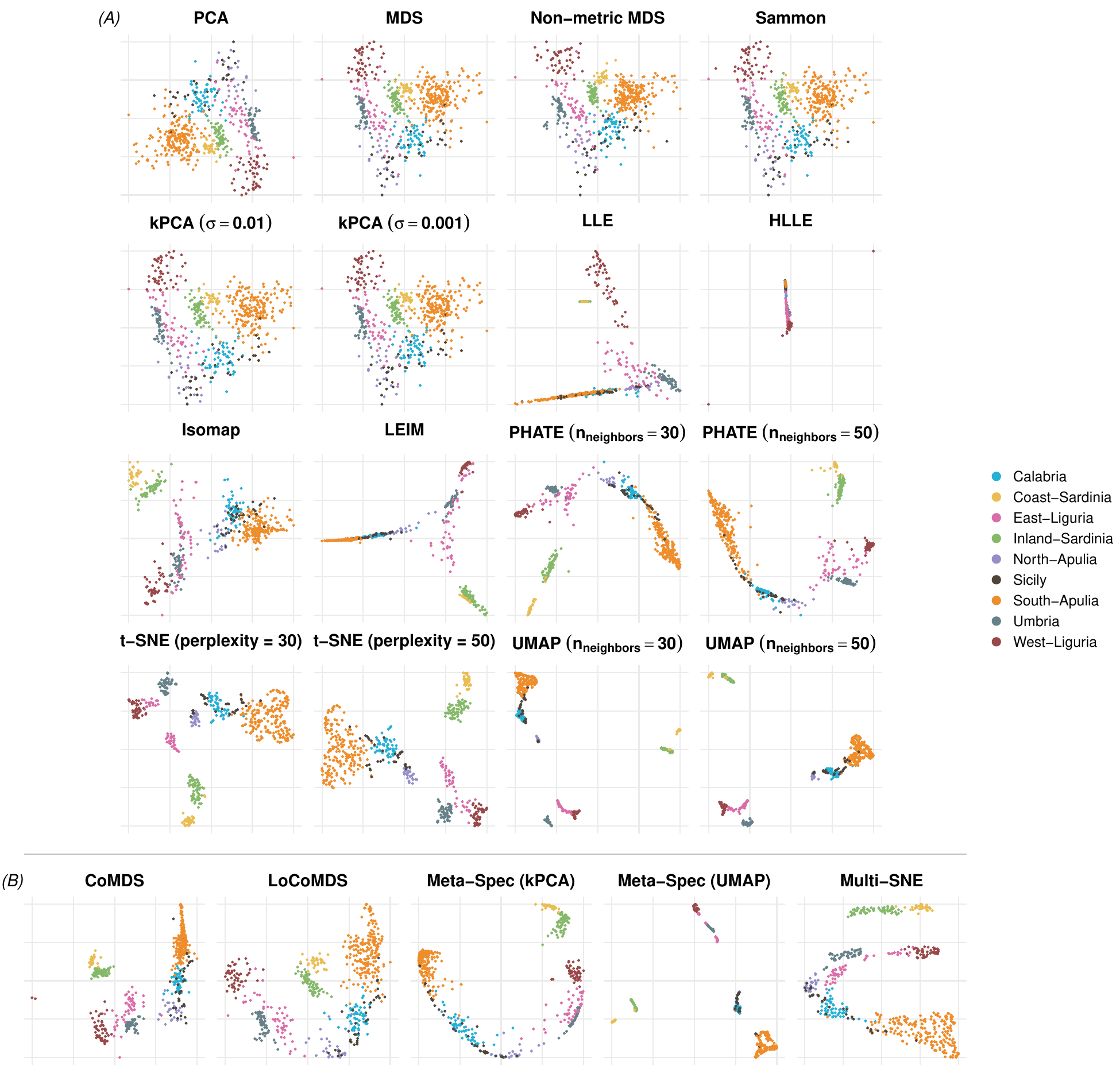}
    \caption{Low-dimensional embeddings, obtained from (A) all 16 input dimension reduction methods and (B) all consensus dimension reduction methods under consideration, applied to the \textbf{olive oil data}.}
    \label{fig:full_olive}
\end{figure}

\begin{figure}
    \centering
    \includegraphics[width=1.0\linewidth]{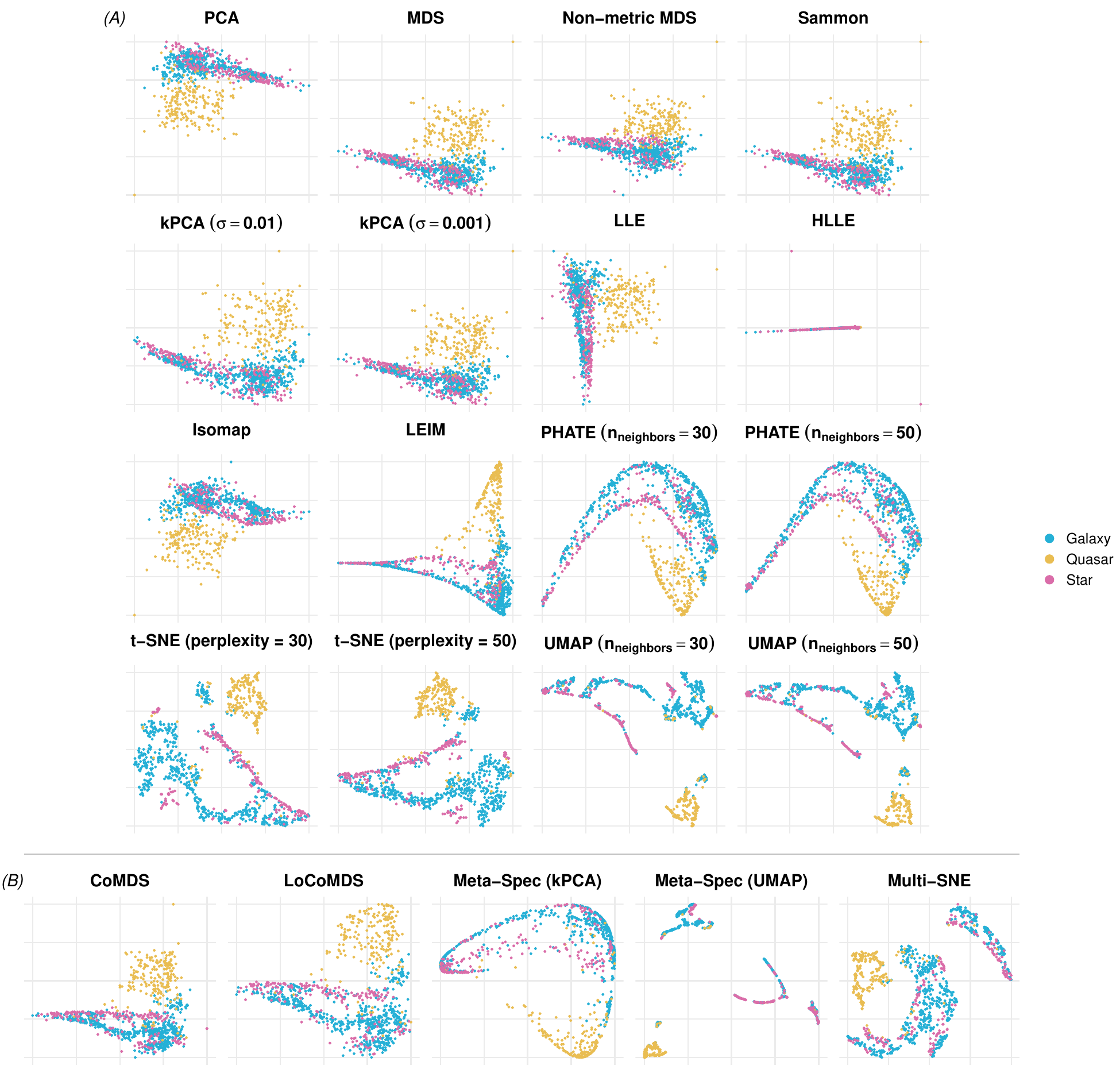}
    \caption{Low-dimensional embeddings, obtained from (A) all 16 input dimension reduction methods and (B) all consensus dimension reduction methods under consideration, applied to the \textbf{star data}.}
    \label{fig:full_star}
\end{figure}

\begin{figure}
    \centering
    \includegraphics[width=1.0\linewidth]{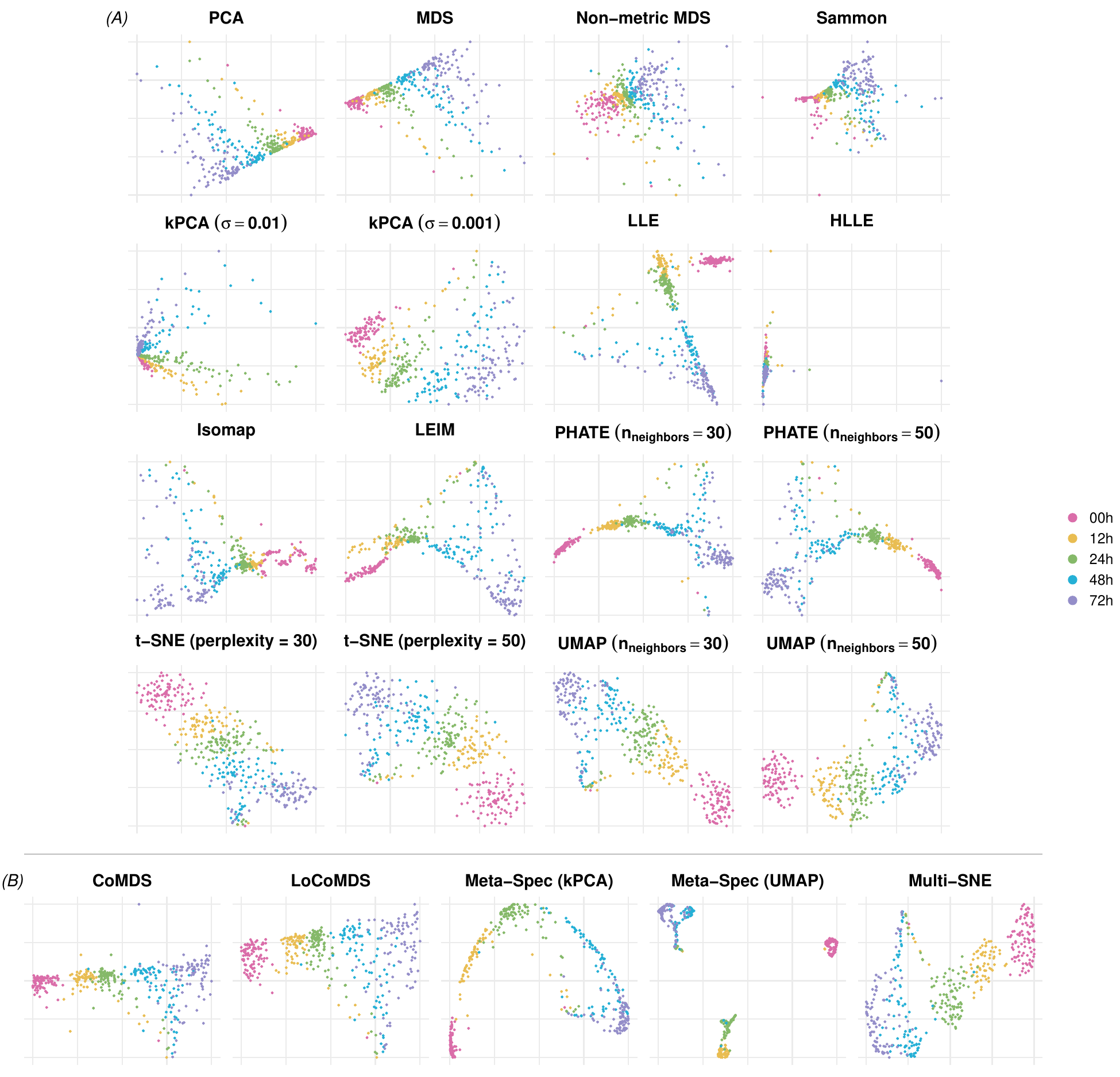}
    \caption{Low-dimensional embeddings, obtained from (A) all 16 input dimension reduction methods and (B) all consensus dimension reduction methods under consideration, applied to the \textbf{trajectory data}.}
    \label{fig:full_trajectory}
\end{figure}

\begin{figure}
    \centering
    \includegraphics[width=1.0\linewidth]{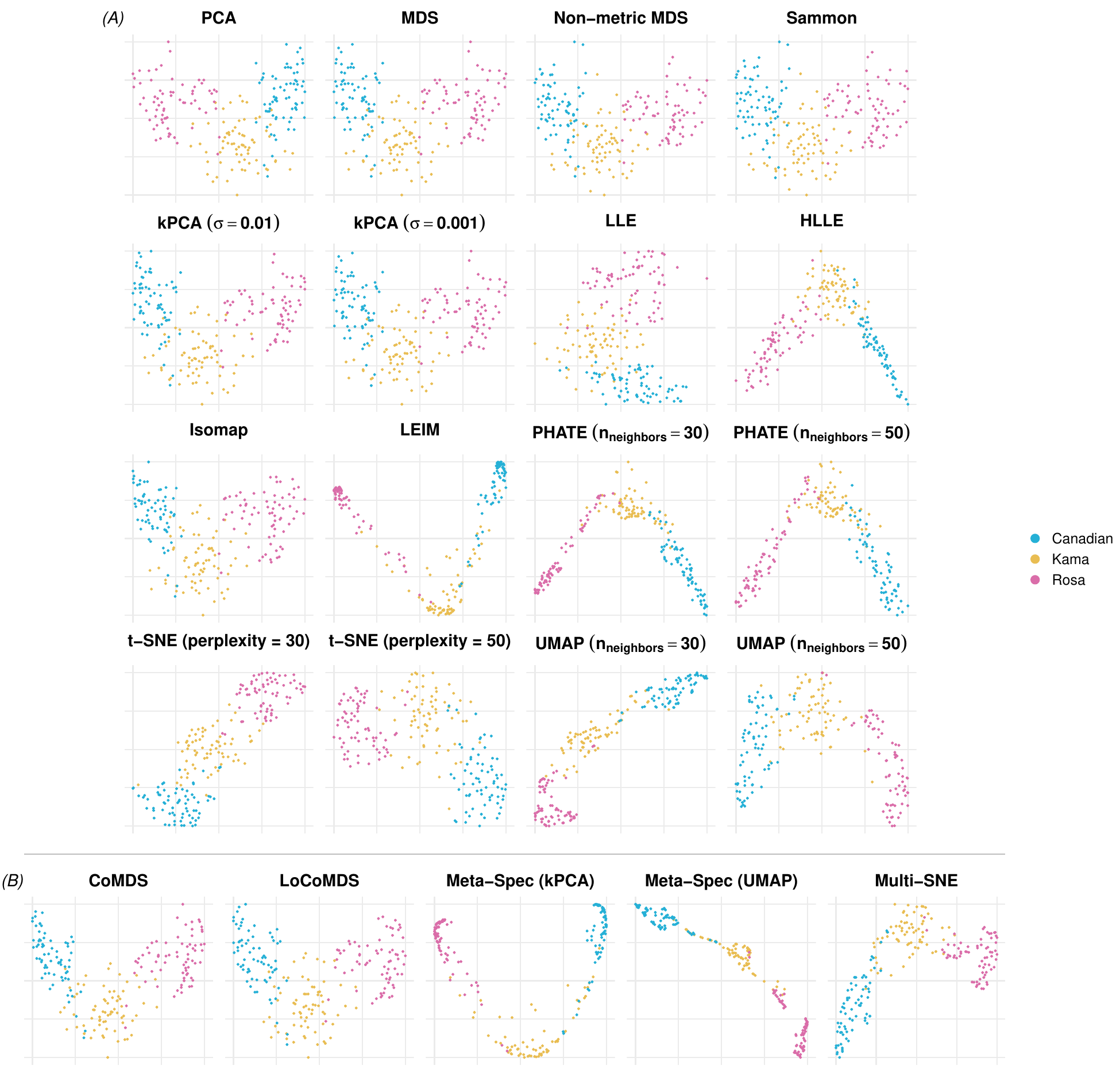}
    \caption{Low-dimensional embeddings, obtained from (A) all 16 input dimension reduction methods and (B) all consensus dimension reduction methods under consideration, applied to the \textbf{wheat data}.}
    \label{fig:full_wheat}
\end{figure}

\begin{figure}
    \centering
    \includegraphics[width=1.0\linewidth]{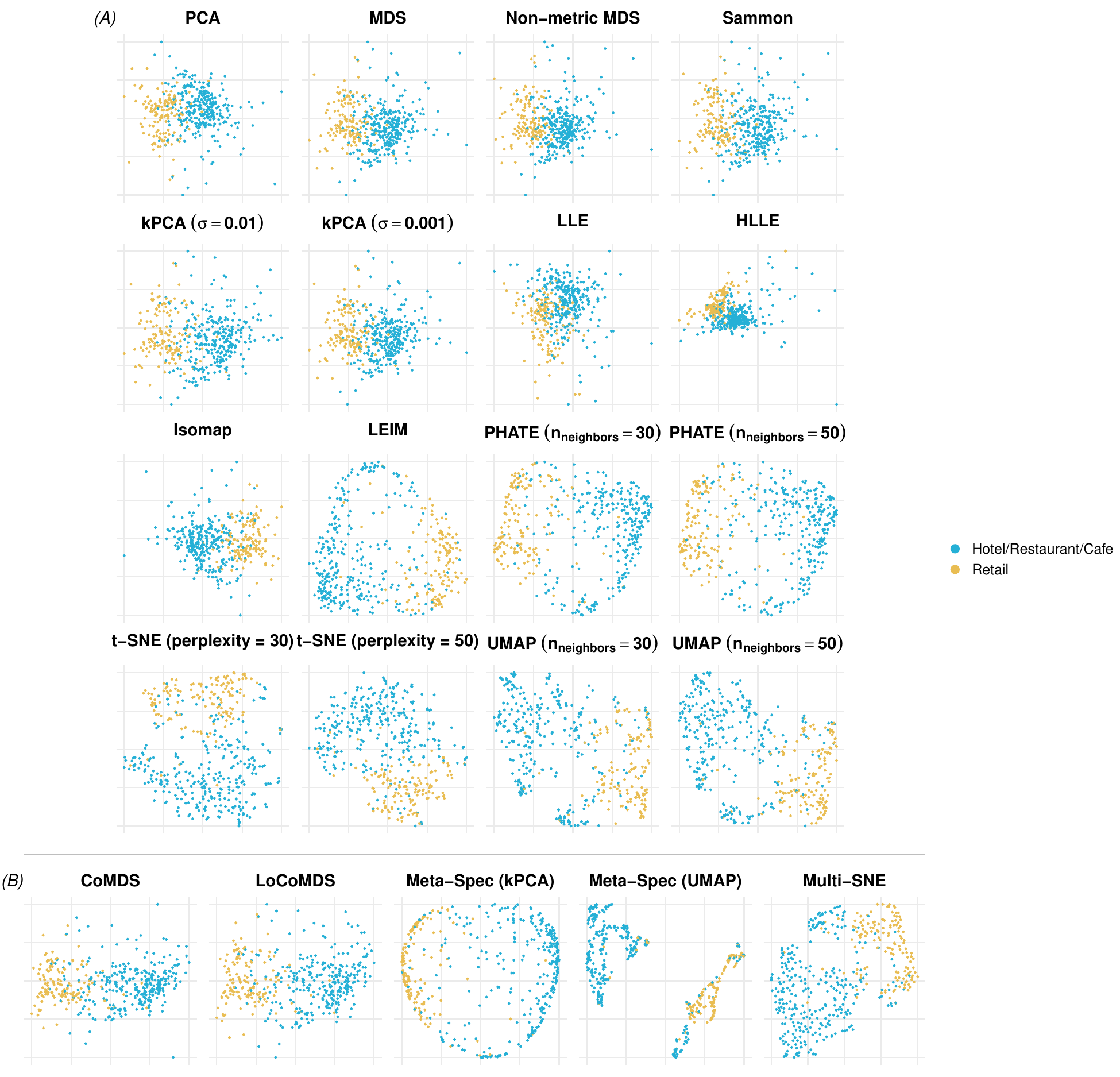}
    \caption{Low-dimensional embeddings, obtained from (A) all 16 input dimension reduction methods and (B) all consensus dimension reduction methods under consideration, applied to the \textbf{wholesale data}.}
    \label{fig:full_wholesale}
\end{figure}


\end{document}